\documentclass[twocolumn]{aastex63} 
\usepackage[version=4]{mhchem}
\usepackage{soul}

\shortauthors{Tsai et al.}
\graphicspath{{./}{figs/}}

\begin{document}

\title{Inferring Shallow Surfaces on sub-Neptune Exoplanets with JWST}
\author[0000-0002-8163-4608]{Shang-Min Tsai}
\affiliation{Atmospheric, Oceanic and Planetary Physics, Department of Physics, University of Oxford, UK}

\author[0000-0001-5271-0635]{Hamish Innes}
\affiliation{Atmospheric, Oceanic and Planetary Physics, Department of Physics, University of Oxford, UK}

\author[0000-0002-3286-7683]{Tim Lichtenberg}
\affiliation{Atmospheric, Oceanic and Planetary Physics, Department of Physics, University of Oxford, UK}

\author[0000-0003-4844-9838]{Jake Taylor}
\affiliation{Department of Physics and Institute for Research on Exoplanets, Universit\'e de Montr\'eal, CA}

\author[0000-0002-2110-6694]{Matej Malik}
\affiliation{Department of Astronomy, University of Maryland, College Park, USA} 

\author[0000-0002-4552-4559]{Katy Chubb}
\affiliation{Centre for Exoplanet Science, University of St Andrews, North Haugh, St Andrews, UK} 
\affiliation{Stellar Astrophysics Centre, Deparment of Physics and Astronomy, Aarhus University, Ny Munkegade 120, Denmark} 

\author[0000-0002-5887-1197]{Raymond Pierrehumbert}
\affiliation{Atmospheric, Oceanic and Planetary Physics, Department of Physics, University of Oxford, UK}


\begin{abstract}
Planets smaller than Neptune and larger than Earth make up the majority of the discovered exoplanets. Those with \ce{H2}-rich atmospheres are prime targets for atmospheric characterization. The transition between the two main classes, super-Earths and sub-Neptunes, is not clearly understood as the rocky surface is likely not accessible to observations. Tracking several trace gases (specifically the loss of ammonia (\ce{NH3}) and hydrogen cyanide (HCN)) has been proposed as a proxy for the presence of a shallow surface. In this work, we revisit the proposed mechanism of nitrogen conversion in detail and find its timescale on the order of a million years. \ce{NH3} exhibits dual paths converting to \ce{N2} or HCN, depending on the UV radiation of the star and the stage of the system. In addition, methanol (\ce{CH3OH}) is identified as a robust and complementary proxy for a shallow surface. We follow the fiducial example of K2-18b with a 2D photochemical model (VULCAN) on an equatorial plane. We find a fairly uniform composition distribution below 0.1 mbar controlled by the dayside, as a result of slow chemical evolution. \ce{NH3} and \ce{CH3OH} are concluded to be the most unambiguous proxies to infer surfaces on sub-Neptunes in the era of the James Webb Space Telescope (JWST).

\vspace{0.6cm}
\end{abstract}
\section{Introduction} \label{sec:intro}
Sub-Neptune-sized planets (R$_p$ $\sim$ 1.6--3.5 R$_{\oplus}$) constitute the main population of exoplanets we have discovered \citep{Hsu2019}. Yet their formation \citep{Bean2021}, atmospheric composition \citep{Moses2013}, and interior structure \citep{Madhu2020,Aguichine2021} are not well understood. A recurring question is whether these planets are close to scaled-up terrestrial planets or scaled-down ice giants. The mass-radius relation, when available, provides the first assessment. However, the bulk density can be highly degenerate from different mass fractions of the \ce{H2}/He envelope and iron/rocky core. Particularly the low-density (0.25 $\rho_{\oplus}$ -- 0.75 $\rho_{\oplus}$) planets can either consist of a dense core enclosed by a massive \ce{H2}/He envelope or a lighter core (dominated by rocky material and/or water) with a shallow \ce{H2}-rich atmosphere.

Atmospheric observations provide potential diagnostics to distinguish the above two classes. \cite{Yu2021} suggest that atmospheric chemistry can be utilized to infer the pressure level of the atmosphere-interior interface. When the atmosphere is thin (e.g., $<$ 10 bar), photochemically unstable gases (e.g., ammonia (\ce{NH3}) and methane (\ce{CH4})) would decline without being reformed through thermochemistry in the deep atmosphere. \cite{Hu2021} focus on the distinctive features of an ocean planet with a thin atmosphere where solubility equilibrium is expected. The temperate sub-Neptune K2-18b with a recent water detection \citep{Benneke2019,Tsiaras2019} is applied in both studies as a fiducial example. While \cite{Yu2021} do not consider atmosphere-surface exchanges, \cite{Hu2021} assume a plausible range of \ce{CO2} concentrations in the initial atmosphere consistent with a massive water ocean. Nevertheless, both conclude that \ce{CH4} in a 1-bar atmosphere is about 10--100 times less abundant than that in a massive atmosphere, independent of the amount of \ce{CO2}. \cite{Yu2021} show that the slow chemical recycling makes \ce{NH3} the most sensitive species and a prominent proxy for the presence of surfaces. However, the amount of \ce{NH3} in a 1-bar atmosphere predicted in both studies differ by orders of magnitude due to different assumptions of initial composition. Lastly, previous studies have commonly relied on 1D models and overlooked the interaction between dayside and nightside.

In this work, we first revisit the destruction mechanism driven by photochemistry that has been suggested to serve as an observational discriminator between shallow and deep atmospheres. We elucidate the timescale and duality of the nitrogen conversion with a quiet and an active M star, and for a range of atmospheric metallicity and vertical mixing. Finally, a key addition to previous work is that we apply a 2D model including day-night transport to reevaluate the viability of utilizing atmospheric chemistry as a proxy for the presence of surfaces.

\section{Methods}
\subsection{1D radiative transfer model}
To facilitate comparison with previous work, we take K2-18b as a test case for identifying the surface underneath a small \ce{H2} atmosphere. The pressure-temperature ($P$--$T$) profiles of K2-18b in dry radiative-convective equilibrium are computed by the radiative transfer model HELIOS \citep{Malik2019,Malik2019b}. We assume a moderate 100 times solar metallicity (same as \cite{Yu2021}) and an internal heat with $T_{\textrm{int}}$ = 30 K \citep{Lopez2013} with uniform heat redistribution. The $P$--$T$ profile is not self-consistently evolving with chemistry but simply fixed as input to the photochemical model. We discuss the radiative effects of the surface in Section \ref{sec:surface} and the radiative feedback of disequilibrium chemistry in Section \ref{sec:discussion}.

\subsection{1D photochemical model}
The atmospheric composition is computed using the photochemical kinetics model VULCAN \citep{tsai17,Tsai2021}. The model treats photolysis, thermochemical kinetics, and vertical mixing. VULCAN has been applied to a variety of planetary atmospheres, including hot-Jupiters, sub-Neptunes, and Earth \citep[e.g.][]{Zilinskas2020,Tsai2021}. We consider species containing N, C, O, H for this study. The large-scale mixing process is parameterized by the eddy diffusion using the expression as a function of pressure in bar ($P_{\textrm{bar}}$) \citep{Tsai2021} in our nominal model
\begin{equation}\label{eq:Kzz}
K_{\textrm{zz}} = 10^5 \left(\frac{1 \textrm{bar}}{P_{\textrm{bar}}}\right)^{0.4} \quad (\textrm{cm}^2/\textrm{s}).   
\end{equation}

We further test the sensitive to eddy diffusion and the effects of surface sinks of ammonia, as discussed in Section \ref{sec:Kzz} and \ref{sec:discussion}.


\subsection{2D photochemical model}\label{sec:method-2D}
The day-night circulation is expected to be crucial in regulating the compositional variation across a tidally-locked planet \citep[e.g.][]{Agundez2014,Drummond2020,Feng2020,Joost2021}. We run the 2D photochemical model VULCAN on a pressure-longitude grid to account for the day-night transport on a meridionally averaged equatorial plane. The 2D model solves the continuity equations including horizontal transport
\begin{equation}
\frac{\partial n_i}{\partial t} = {\cal P}_i - {\cal L}_i - \frac{\partial \phi_{i,z}}{\partial z} - \frac{\partial \phi_{i,x}}{\partial x},
\label{eq:master}
\end{equation}
where $n_i$ is the number density (cm$^{-3}$) of species $i$, ${\cal P}_i$ and ${\cal L}_i$ are the chemical production and loss rates (cm$^{-3}$ s$^{-1}$) of species $i$, and $\phi_{i,z}$, $\phi_{i,x}$ are the vertical and horizontal transport flux, respectively. The construction and benchmarks of 2D VULCAN are discussed in detail in a follow-up paper (in prep). The 3D global circulation model (GCM) Exo-FMS with double grey radiation scheme is utilized to set up the input for the 2D photochemical model. Exo-FMS has previously been used in studies of both terrestrial exoplanet \citep{Pierrehumbert2016, Hammond2017,Hammond2018,Ray2019} and gas giant \citep{Lee2020,Lee2021} atmospheres. The temperature and wind fields are averaged over the board equatorial zone across 45$^{\circ}$, where the circulation is dominated by a zonal jet and well represented by a 2D framework. The equatorial region is then divided into four quarters by longitude: dayside (325$^{\circ}$ -- 45$^{\circ}$), morning limb (45$^{\circ}$ -- 135$^{\circ}$), nightside (135$^{\circ}$ -- 225$^{\circ}$), and evening limb (225$^{\circ}$ -- 325$^{\circ}$). The averaged zenith angles for the four quarters are 32$^{\circ}$, 72$^{\circ}$, 0$^{\circ}$, 72$^{\circ}$, respectively. The root-mean-square of the vertical wind velocity ($w_{\textrm{rms}}$) is converted to eddy diffusion by the mixing length theory $K_{\textrm{zz}} = w_{\textrm{rms}} H$, where $H$ is the scale height as the characteristic length scale.  
\subsection{Transmission spectrum modeling}
 We use the radiative transfer and retrieval framework NEMESIS \citep{Irwin2008} to produce the transmission spectra. The forward models were computed using the correlated-$k$ technique, with the $k$-tables being computed using the methodology of \citet{2021Chubb}. Specifically, the sources of the opacity data for the molecules of interest are: NH$_3$ \citep{NH3_opacity}, CO \citep{CO_opacity},  CO$_2$ \citep{CO2_opacity}, CH$_3$OH\footnote{The opacity of \ce{CH3OH} is only available at room temperature but the temperature in the region sensitive to transmission ($P <$ 0.1 bar) is also around 300 K.} \citep{CH3OH_opacity_1}. 
 
\section{1D Results: Slow Loss of Ammonia}\label{sec:1D_results}
\subsection{Minimal surface effects on the thermal structure}\label{sec:surface}
Before looking into the compositional diagnostics, we first discuss how the existence of surfaces might impact the thermal structure. The radiative effect of the surface is neglected in \cite{Yu2021} where the temperature profile is simply truncated at different pressure levels. In practice, the surface partially absorbs stellar irradiation and re-emits back to the atmosphere. A canonical example is that of Earth, where roughly half of the solar radiation is absorbed by the surface but the atmosphere is mostly opaque to infrared radiation. This potentially makes the temperature higher than that of a ``surface-less'' atmosphere at the same pressure level, which can in turn alter the composition and the ability to recycle atmospheric species. 

\begin{figure}[ht!]
\plotone{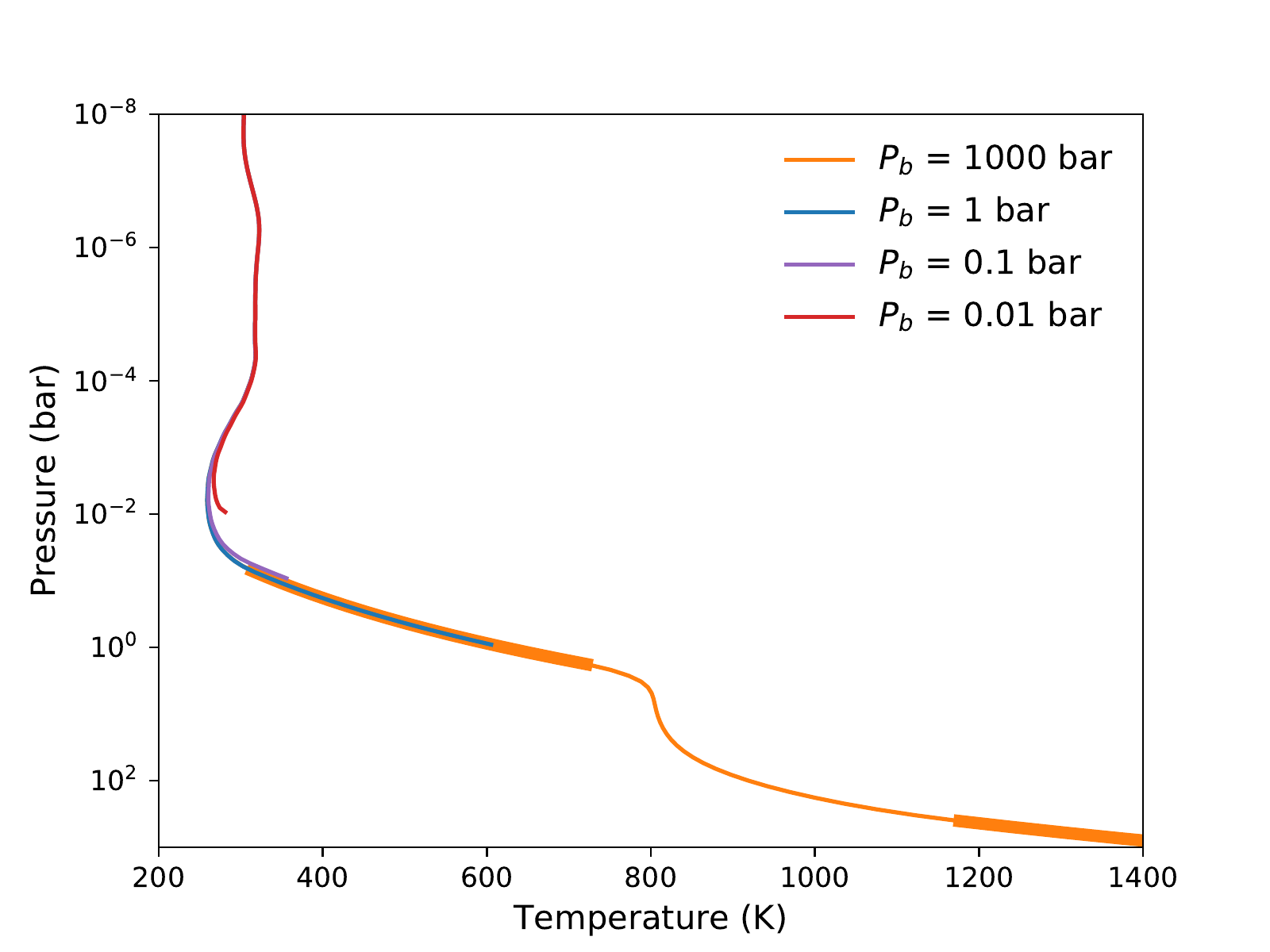}
\caption{The pressure--temperature profiles of K2-18b with various surface pressure levels ($P_b$). Thickened lines indicate the convective regions. A solid surface appears to have minimal effect on the $P$--$T$ profiles once the atmosphere is sufficiently opaque.
\label{fig:TPs}}
\end{figure}

We test the thermal effects of the surface using 1D radiative-convective calculations with a generic surface placed at different pressure levels. We assume a surface albedo of 0.1 but find no differences in the range of 0.1--0.3 (corresponding to the albedo range of land areas on Earth)\footnote{Note that we assume a wavelength-independent surface albedo, which likely underestimates the impact of the albedo on the surface temperature.}. The resulting temperature profiles for 100 times solar metallicity are depicted in Figure \ref{fig:TPs}. The convective zone extends from a few bar to 0.1 bar and the temperature is set by the dry adiabat. The presence of a surface turns out to have negligible thermal effects in this region since the atmosphere absorbs most of the stellar irradiation. Only for surface pressures around 0.01 bar, the greenhouse warming near the surface starts to be notable, because the atmosphere is optically thin toward stellar radiation while being thermally opaque in this region. In all cases, our test validates the approach of directly truncating the model for shallow atmospheres with surface pressure down to about 1 bar (also see \cite{May2020} for the global heat transport).

\subsection{Nitrogen conversion for a quiet M star and an active M star}
Photochemically active gases, including \ce{NH3}, HCN, and \ce{CH4}, are lost in the absence of a deep atmosphere where themochemistry operates fast and fully recycles them back. Comparing the results of a surface at 1 bar to the deep/no surface case, \cite{Yu2021} find that \ce{NH3} is decreased by about 5 orders of magnitude while HCN is decreased by about 3 orders of magnitude. In this section,  we will address the timescale of nitrogen conversion and overall compositional evolution driven by photochemistry. It is relevant to consider the host star at different stages when the timescale of chemical destruction is comparable to the timescale of stellar evolution. We apply the synthetic M2 star at the age of 45 Myr and 5 Gyr from HAZMAT \citep{Peacock2020} for the stellar UV spectra, to represent a young, active and an old, quiet M star\footnote{\cite{Yu2021} consider a young, active M star with the UV flux compiled from several stars.}.

\begin{figure*}[ht!]
\begin{center}
\includegraphics[width=\columnwidth]{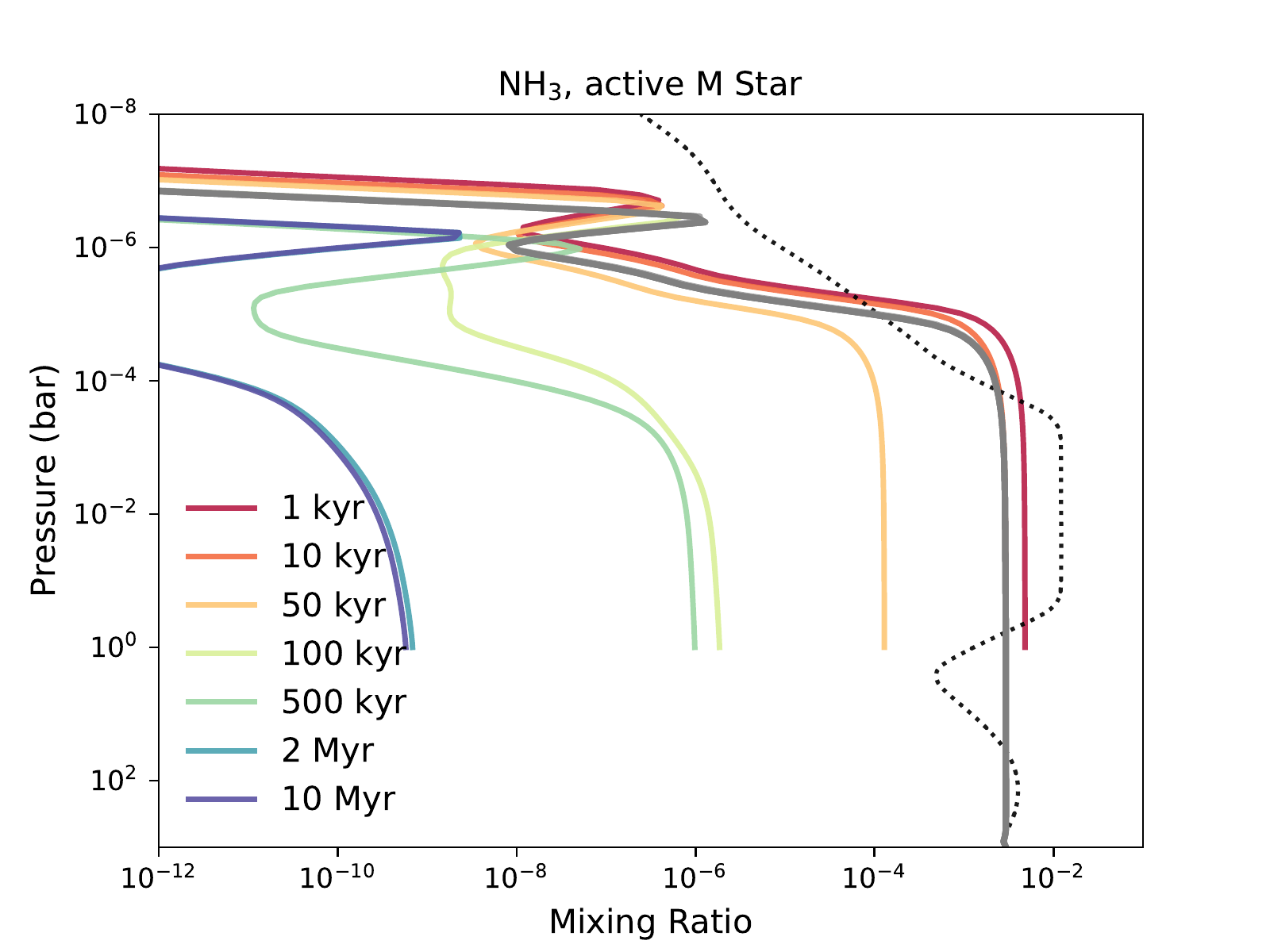}
\includegraphics[width=\columnwidth]{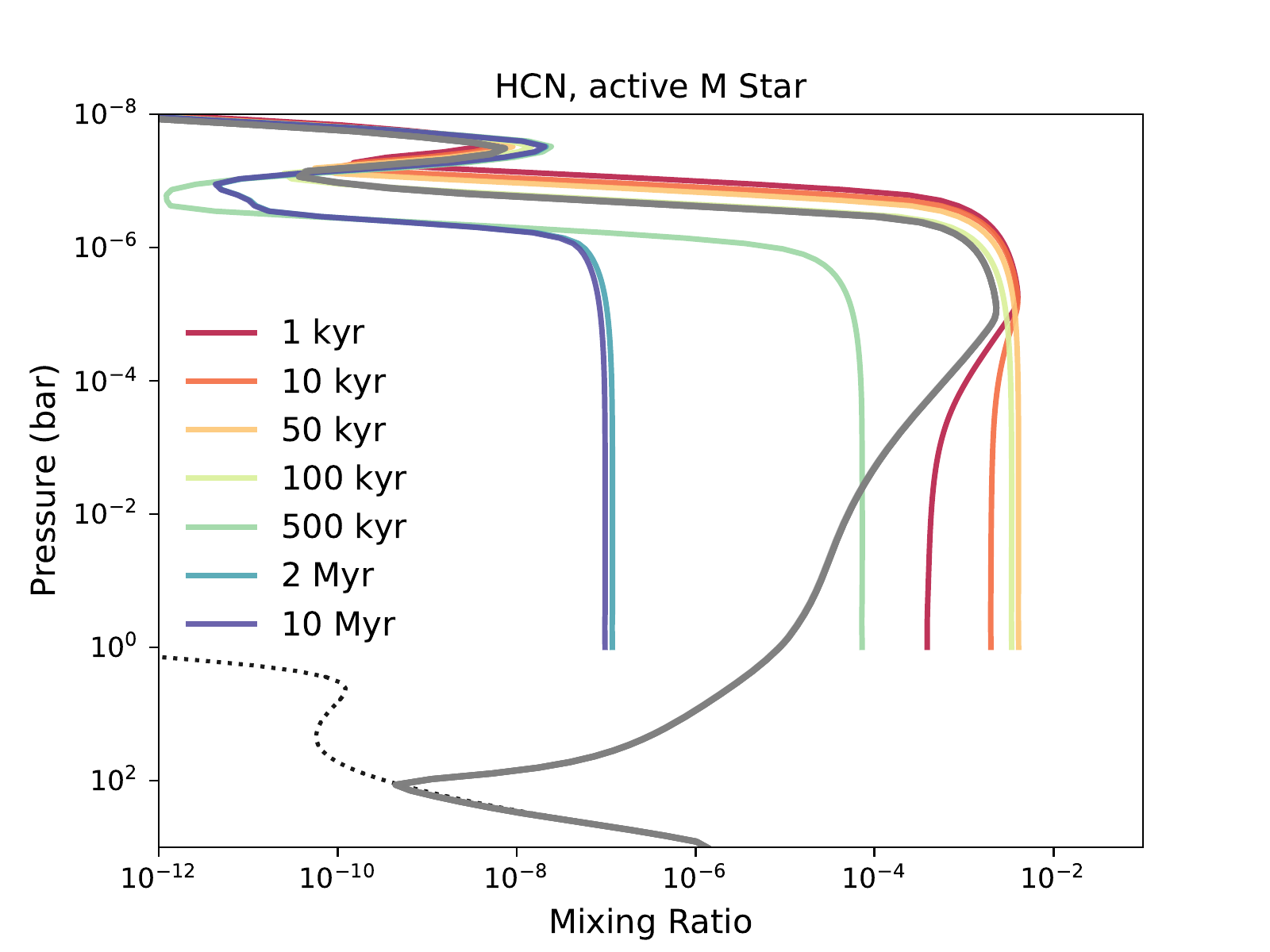}
\includegraphics[width=\columnwidth]{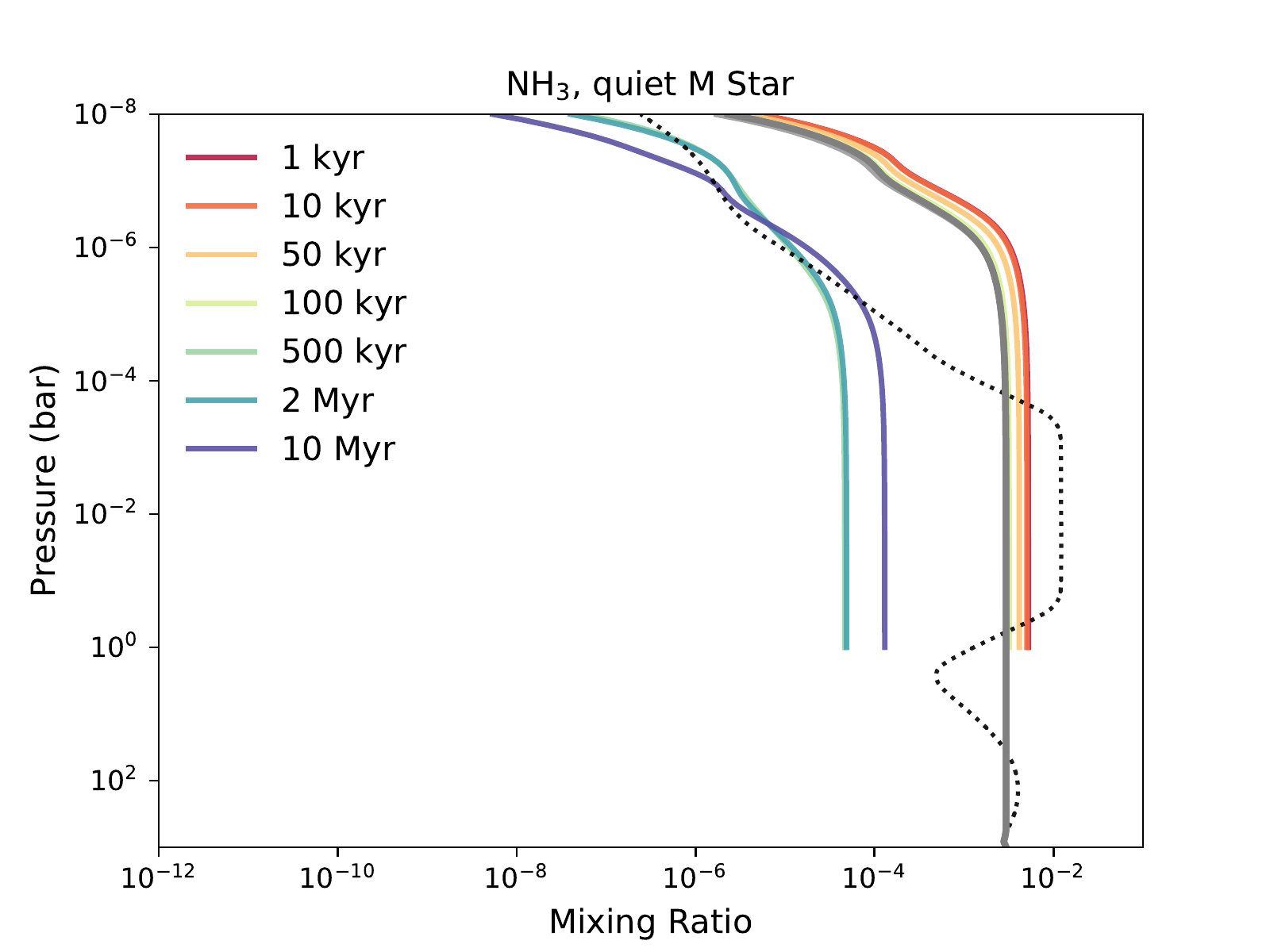}
\includegraphics[width=\columnwidth]{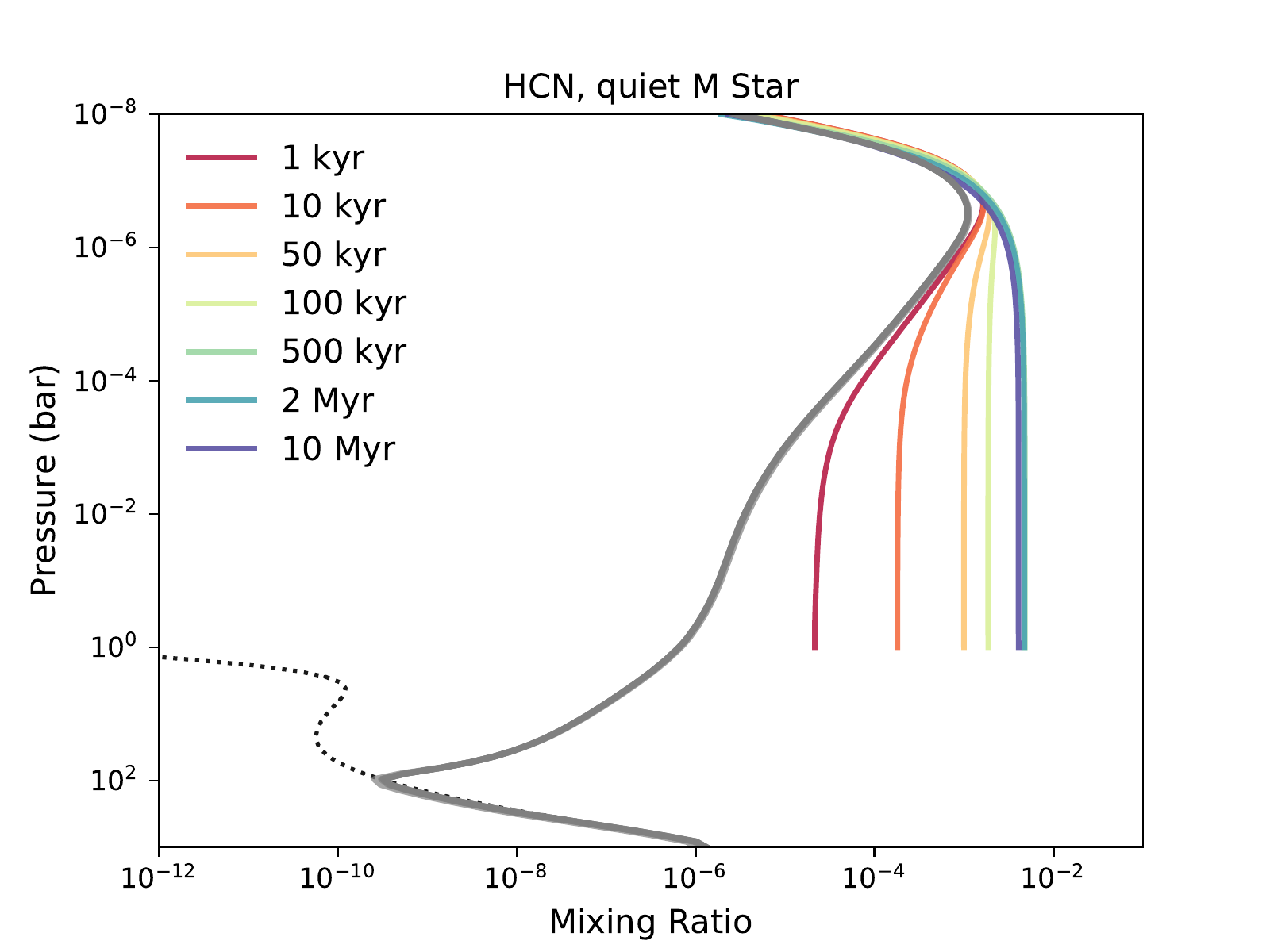}
\end{center}
\caption{The evolution of the mixing ratio profiles of \ce{NH3} (left) and HCN (right) with a 1-bar surface compared to a deep, 1000-bar atmosphere for an active (top) and a quiet (bottom) M star. The snapshot abundances at different time in the 1-bar surface model are color coded from red to blue. The abundances in the deep model
do not evolve after 1 kyr because of the efficient thermochemistry in the deep atmosphere and are shown as solid grey lines. The black dotted curves indicate the initial equilibrium abundances. The upper panels are for an active M2 star at the age of 45 Myr and the lower panels for a quite M2 star at the age of 5 Gyr.}
\label{fig:nh3_1D_time}
\end{figure*}

Figure \ref{fig:nh3_1D_time} shows snapshots of \ce{NH3} and HCN abundances at different times with a 1-bar surface compared with a deep, 1000-bar surface, for both a quiet M star and an active M star\footnote{The complete abundance profiles are available in the supplementary materials}. All models start from initial abundances in chemical equilibrium. We first confirm that while recycling is enabled in the deep surface scenario, \ce{NH3} and HCN settle to a steady state within 1000 years. In the 1-bar atmosphere, \ce{NH3} mostly drops off between 10 kyr and 1 Myr to a final mixing ratio several orders of magnitudes lower, qualitatively matching the decrease of \ce{NH3} in \cite{Yu2021}. However, as \ce{NH3} is gradually lost, HCN {\it does not} readily follow \ce{NH3}. With a quiet M star, HCN actually increases to a uniform abundance close to 1$\%$ in the period between 0.1 and 10 Myr before gradually declining after 10 Myr.

\begin{figure*}[ht!]
\begin{center}
\vspace{-0.75cm}
\gridline{\fig{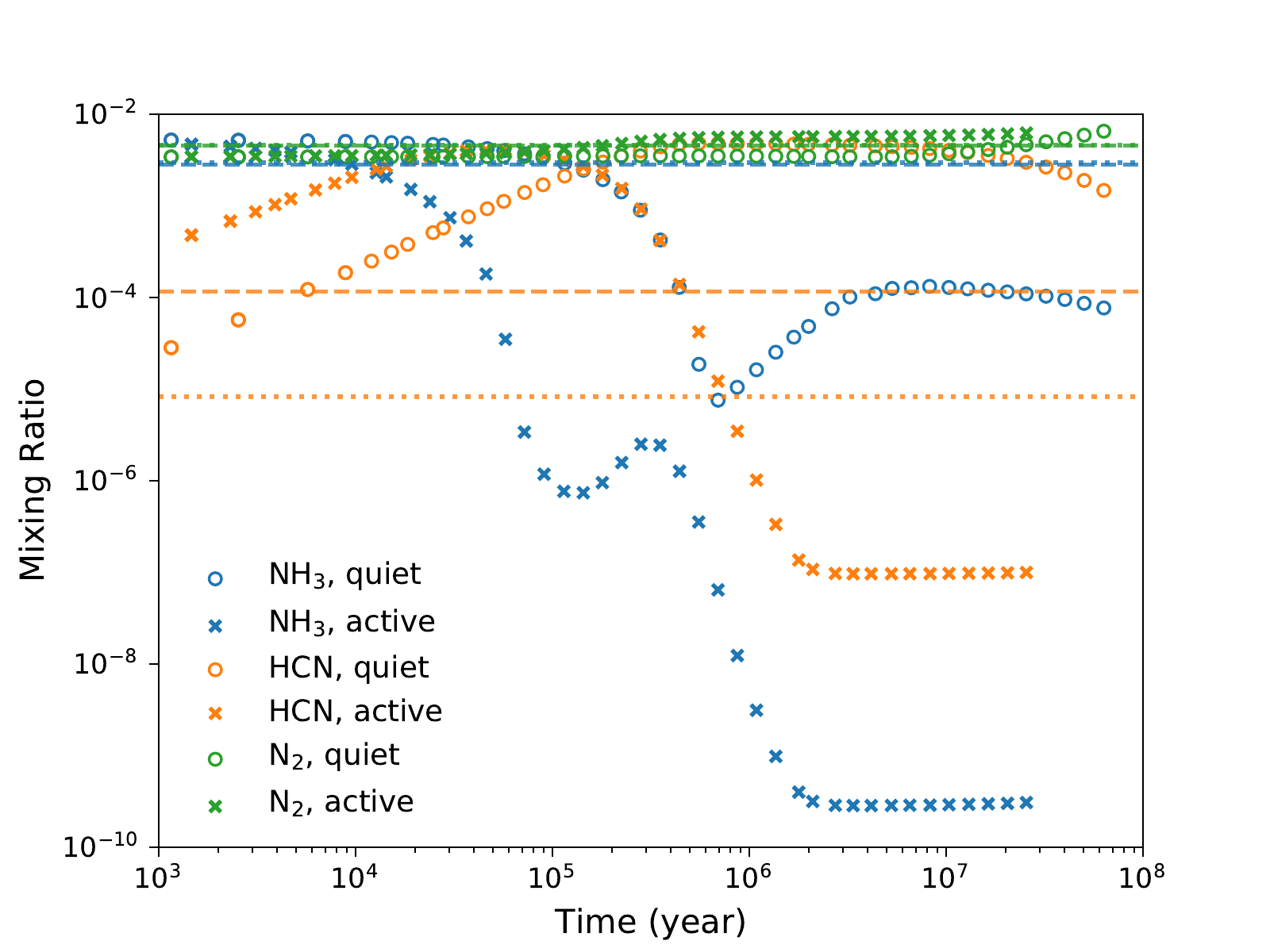}{\columnwidth}{(a)}
\fig{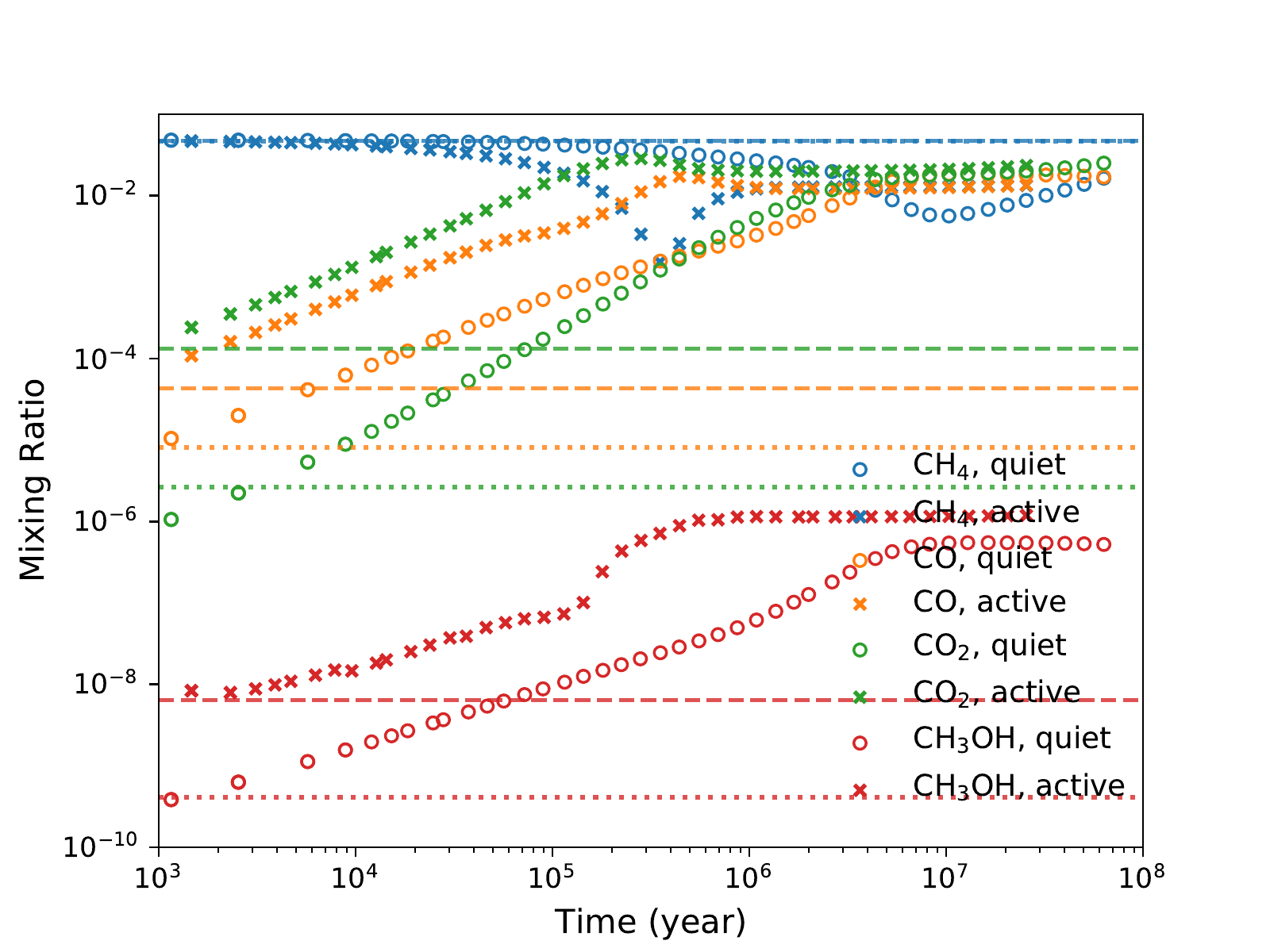}{\columnwidth}{(b)}}
\vspace{-0.5cm}
\gridline{\fig{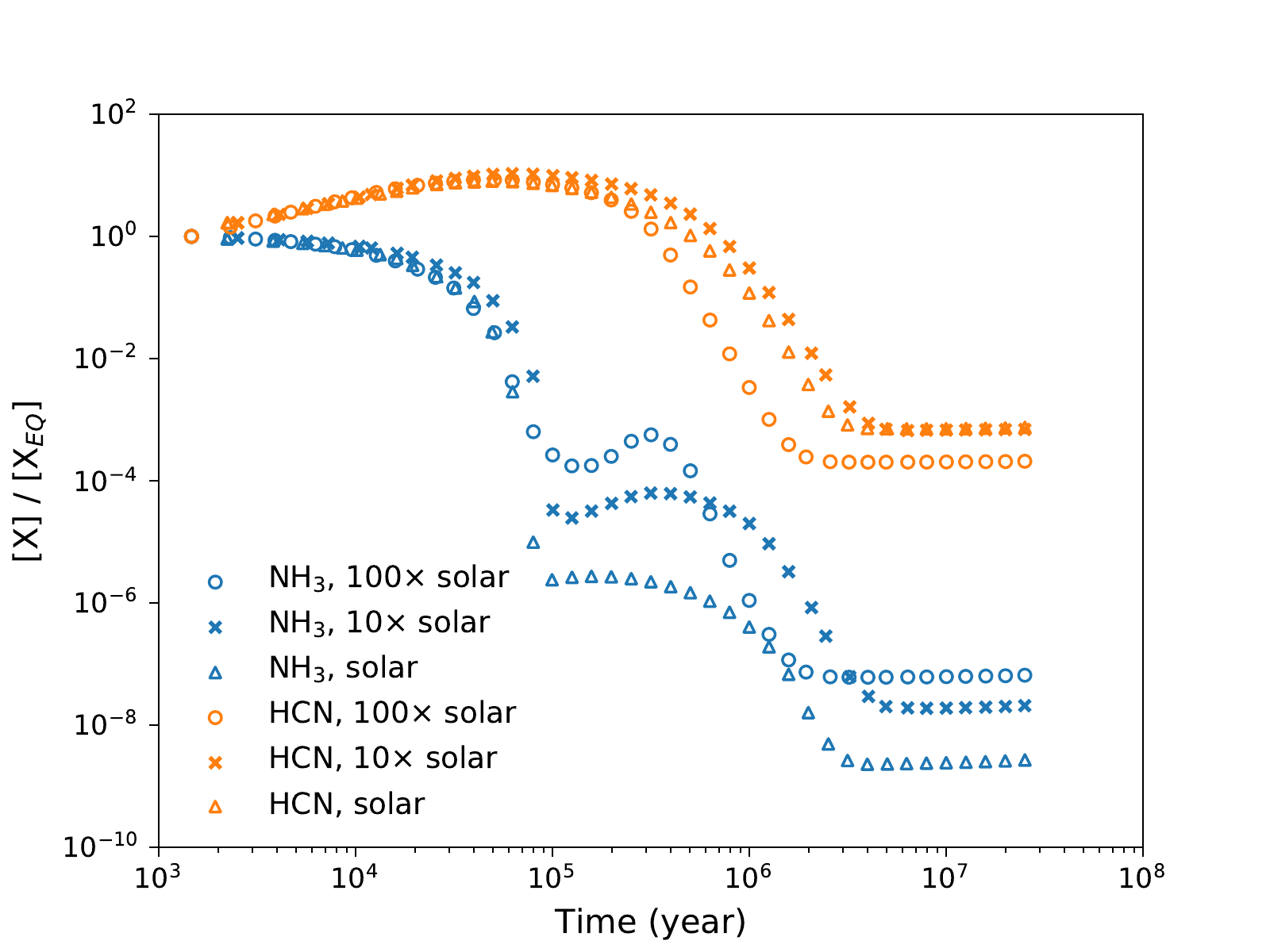}{\columnwidth}{(c)} 
\fig{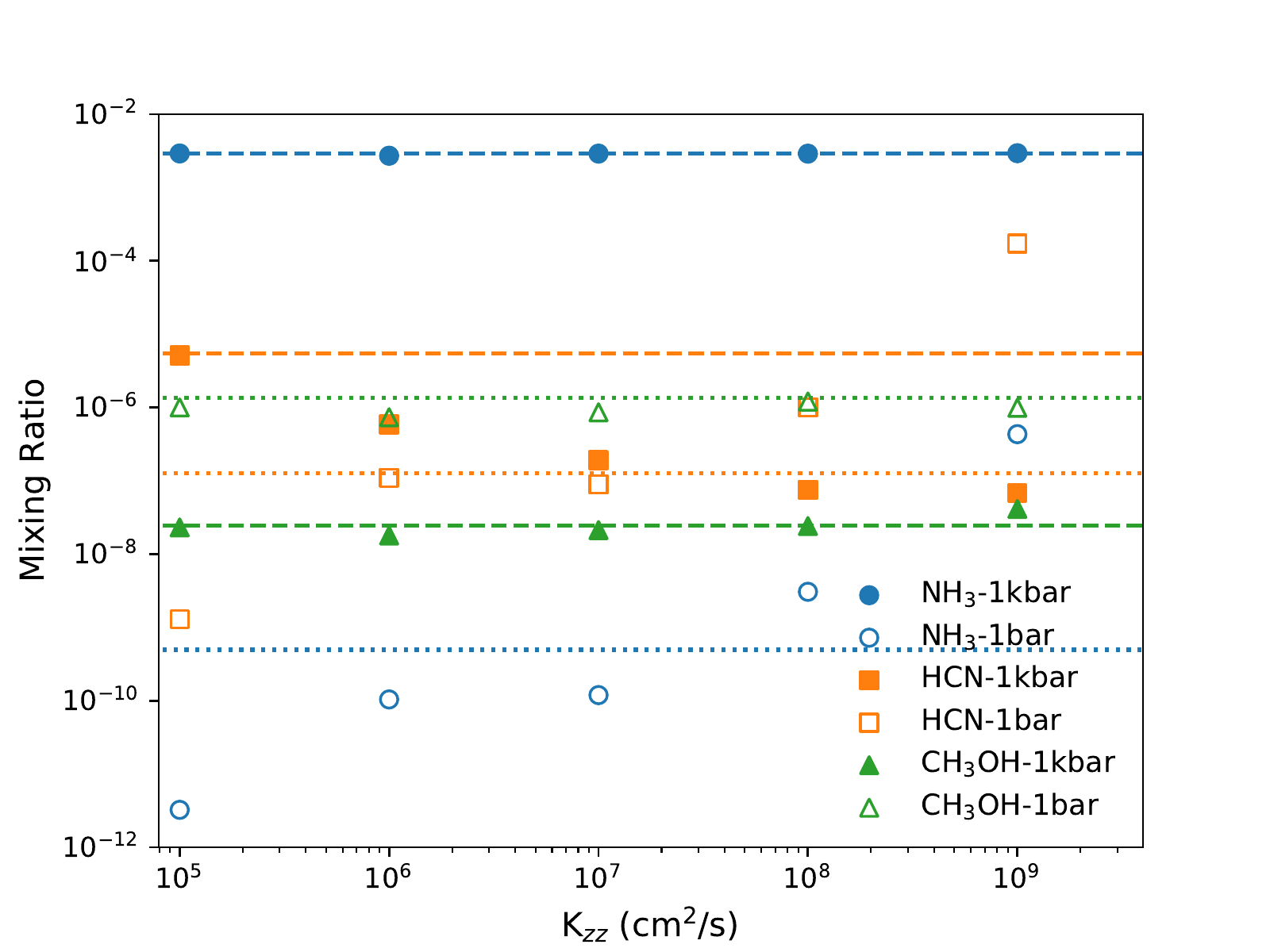}{\columnwidth}{(d)} }
\end{center}
\vspace{-0.3cm}
\caption{(a) The evolution of the column-averaged mixing ratios of \ce{NH3} and HCN in an atmosphere with a 1-bar surface for a quiet and an active M star, where the dotted (quiet) and dashed (active) lines indicate the mixing ratios from the 1000-bar models for comparison. The column average is taken in the main observable part between 1 and 0.1 mbar where their volume mixing ratios are already close to uniform (Figure \ref{fig:nh3_1D_time}). (b) Same as (a) but for \ce{CH4}, CO, \ce{CO2} and \ce{CH3OH}. (c) The evolution of the ratios between the abundances at a given time and their equilibrium abundances with an active M star for different atmospheric metallicities. (d) The final column-averaged mixing ratios of \ce{NH3}, HCN, and \ce{CH3OH} at 100 Myr with an active M star and a 1-bar surface (filled symbols) or 1000-bar surface (open symbols) for different values of uniform eddy diffusion. The dashed and dotted lines indicate the mixing ratios from the nominal 1000-bar and 1-bar models, respectively, where the eddy diffusion profile is given by Equation (\ref{eq:Kzz}).}
\label{fig:mix_time}
\end{figure*}

The timescale of nitrogen conversion given by the evolution of major nitrogen species is further captured in Figure \ref{fig:mix_time} (a). Overall, we find that nitrogen conversion takes place between 10 kyr and 1 Myr for an active M star and is much slower for a quiet M star. \ce{NH3} is evidently converted to \ce{N2} for an active M star, but HCN can serve as a transient nitrogen pool in the case of a quiet M star.

\subsubsection{The two conversion paths of \ce{NH3}}
HCN is produced in the upper atmosphere initiated by \ce{NH3} photolysis:
\begin{eqnarray}
\begin{aligned}
\ce{NH3 &->[h\nu] NH2 + H}\\
\ce{CH4 + H &-> CH3 + H2}\\
\ce{NH2 + CH3 &->[M] CH3NH2}\\
\ce{CH3NH2 + H &-> CH2NH2 + H2}\\
\ce{CH2NH2 + H &-> CH2NH + H2}\\
\ce{CH2NH + H &-> H2CN + H2}\\
\ce{H2CN + H &-> HCN + H2}\\
2(\ce{H2O &->[h\nu] OH + H})\\
2(\ce{OH + H2 &-> H2O + H})\\
\noalign{\vglue 3pt} 
\hline 
\noalign{\vglue 3pt}
\mbox{net} : \ce{NH3 + CH4 &-> HCN + 3H2}. 
\end{aligned}
\label{re:HCN-path}
\end{eqnarray}
While HCN is subject to photodissociation into cyanide radical (CN) and H, CN can rapidly recycle back to HCN with \ce{CN + H2 -> HCN + H} in an \ce{H2} atmosphere. In fact, the major effective sink of HCN is the OH radical produced by water photolysis. HCN is consumed by OH as C irreversibly goes into CO and N converted into \ce{N2}. The primary scheme for the loss of \ce{NH3} and HCN through oxidation is
\begin{eqnarray}
\begin{aligned} 
\ce{NH3 &->[h\nu] NH2 + H}\\
2(\ce{H2O &->[h\nu] OH + H})\\
\ce{CO + OH &-> CO2 + H}\\
\ce{CO2 &->[h\nu] CO + O}\\
\ce{HCN + OH &-> HNCO + H}\\
\ce{HNCO + H &-> CO + NH2}\\
\ce{NH2 + O &-> HNO + H}\\
\ce{HNO + H &-> NO + H2}\\
2(\ce{H + H &->[M] H2})\\
\ce{NH2 + NO &-> N2 + H2O}\\
\noalign{\vglue 3pt} 
\hline 
\noalign{\vglue 3pt} 
\mbox{net} : \ce{NH3 + HCN + H2O &-> N2 + CO + 3H2}. 
\end{aligned}
\label{re:HCN-45myr}
\end{eqnarray}
The critical difference between a quiet and an active M star is that an active M star produces orders-of-magnitude more OH radicals and hence HCN is scavenged more efficiently after building up. On the other hand, HCN is gradually consumed by OH only after 10 Myr with a quiet M star. We can estimate the overall nitrogen conversion timescale of scheme (\ref{re:HCN-45myr}) with its rate-limiting step \citep{tsai18}
\begin{equation}\label{tau_nh3}
\tau = \frac{[\ce{NH3}]}{k[\ce{NH2}][\ce{NO}]}
\end{equation}
where $k$ is the rate coefficient for the reaction \ce{NH2 + NO -> N2 + H2O}. Taking the initial abundances at 0.1 mbar and 1000 years, the timescale of \ce{NH3} by (\ref{tau_nh3}) is $\sim$ 0.1 Myr for an active M star and $\sim$ 10 Gyr for a quiet M star, consistent with the kinetics results. This is not to be confused with the timescale of \ce{NH3} photodissociation, which operates much faster, about a few hours at the same pressure level.

We conclude this section by reiterating that although the ammonia loss in the shallow-surface case is driven by photochemistry, the process of nitrogen conversion occurs over a timescale of a million years or longer for quiet M stars. It resembles denitrification processes on a geological timescale but only takes place in the atmosphere rather than shifting between reservoirs. 

\subsection{Carbon conversion} 
Most of the carbon is initially bound in the photochemically unstable \ce{CH4} and converted into CO and \ce{CO2} over time. The same evolution of carbon species is shown in Figure \ref{fig:mix_time} (b). First, carbon conversion takes about the same timescale as the nitrogen conversion, $\mathcal{O}$(Myr). Second, the conversion of \ce{CH4} to CO and \ce{CO2} is initiated by the \ce{CH4} photodissociation channel: \ce{CH4 ->[h\nu] CH + H2 + H}. CH reacts with \ce{H2O} to form \ce{H2CO}, which is readily photodissociated into CO. This conversion proceeds faster with an active M star but the final abundances of CO and \ce{CO2} remain close independent of the host star. Third, \ce{CH4} is still continuously evolving after millions of years with a quiet M star, making it ambiguous to compare with the deep-atmosphere abundance as a proxy for surface pressure. Last and most interestingly, methanol (\ce{CH3OH}) is produced as a by-product of carbon conversion. A small part of \ce{H2CO}, instead of being photodissociated, can thermally react with hydrogen to form \ce{CH3O} and \ce{CH3OH}. The mixing ratio of \ce{CH3OH} is increased to $\sim$ 10$^{-6}$, compared to $\lesssim$ 10$^{-9}$ and $\lesssim$ 10$^{-8}$ in the deep surface scenario for a quiet and an active M star, respectively, where \ce{CH3OH} is transported and destroyed in the deep atmosphere.

\subsection{Sensitivity to metallicity}
Sub-Neptunes can in general have a wide range of atmospheric metallicity \citep{Moses2013,Fortney2013}. The equilibrium abundance of \ce{NH3} decreases with increasing metallicity as a result of hydrogen deficiency. To test the sensitivity of \ce{NH3} conversion to metallicity, we further explore lower metallicities with the same $P$--$T$ structure. Figure \ref{fig:mix_time} (c) illustrates the evolution of major nitrogen species for 1$\times$, 10$\times$, and 100$\times$ solar metallicity, with a 1-bar surface and an active M star, where the abundances are normalized to those in equilibrium for a clearer comparison between different metallicities. The final [\ce{NH3}]/[\ce{NH3}]$_\textrm{EQ}$ slightly decreases with smaller metallicity while [HCN]/[HCN]$_\textrm{EQ}$ is almost identical for all metallicities. We find that the trends of \ce{NH3} and HCN destruction remain robust. The $\mathcal{O}$(Myr) timescale of nitrogen conversion holds across different metallicities as well.

\subsection{Sensitivity to vertical mixing}\label{sec:Kzz}
The eddy diffusion profile in our nominal model prescribed by Equation (\ref{eq:Kzz}) is close to that in \cite{Yu2021}. The inverse square root of pressure dependence of $K_{\textrm{zz}}$ is commonly assumed to reflect the turbulent mixing due to gravity wave breaking in stratified atmospheres \citep[e.g.,][]{Lindzen1981,Moses2016}. Since the eddy diffusion cannot be derived from first principles and is often treated as a free parameter, we further test the sensitivity to the strength and structure of vertical mixing by running the 1-bar surface case for a range of uniform $K_{\textrm{zz}}$ profiles from 10$^5$ to 10$^9$ cm$^2$/s. This range of eddy diffusion coefficient is consistent with that derived from the average vertical wind in the GCM of K2-18b (see Figure \ref{fig:GCM}) and other sub-Neptunes in a similar dynamical regime (e.g., \cite{Charnay2015}).

The mixing ratios of \ce{NH3}, HCN, and \ce{CH3OH} for different assumptions of eddy diffusion coefficient are reported in Figure \ref{fig:mix_time} (d). \ce{NH3} in the deep model retains a uniform abundance (Figure \ref{fig:nh3_1D_time}) and does not depend on the eddy diffusion, whereas \ce{NH3} decrease in the shallow model generally correlates with smaller eddy diffusion, since the lower well-mixed region leads to deeper UV penetration and photochemical destruction. Overall, \ce{NH3} always has lower abundances in the shallow model than those in the deep model but the exact ratio, [\ce{NH3}]$_\textrm{1bar}$/[\ce{NH3}]$_\textrm{1kbar}$, depends on the strength and shape of 
$K_{\textrm{zz}}$. On the other hand, \ce{CH3OH} consistently shows little sensitivity to eddy diffusion and about two orders of magnitude increase with respect to the deep model. Conversely, HCN highly depends on eddy diffusion as a result of the balance between photochemical production/destruction in the upper atmosphere and vertical transport. HCN can either raise or lower in the presence of a shallow surface, making it unsuitable as a proxy for the surface level.

\section{2D Results: Globally Uniform Distribution controlled by the dayside}\label{sec:2D}
\begin{figure*}[ht!]
\begin{center}
\gridline{\fig{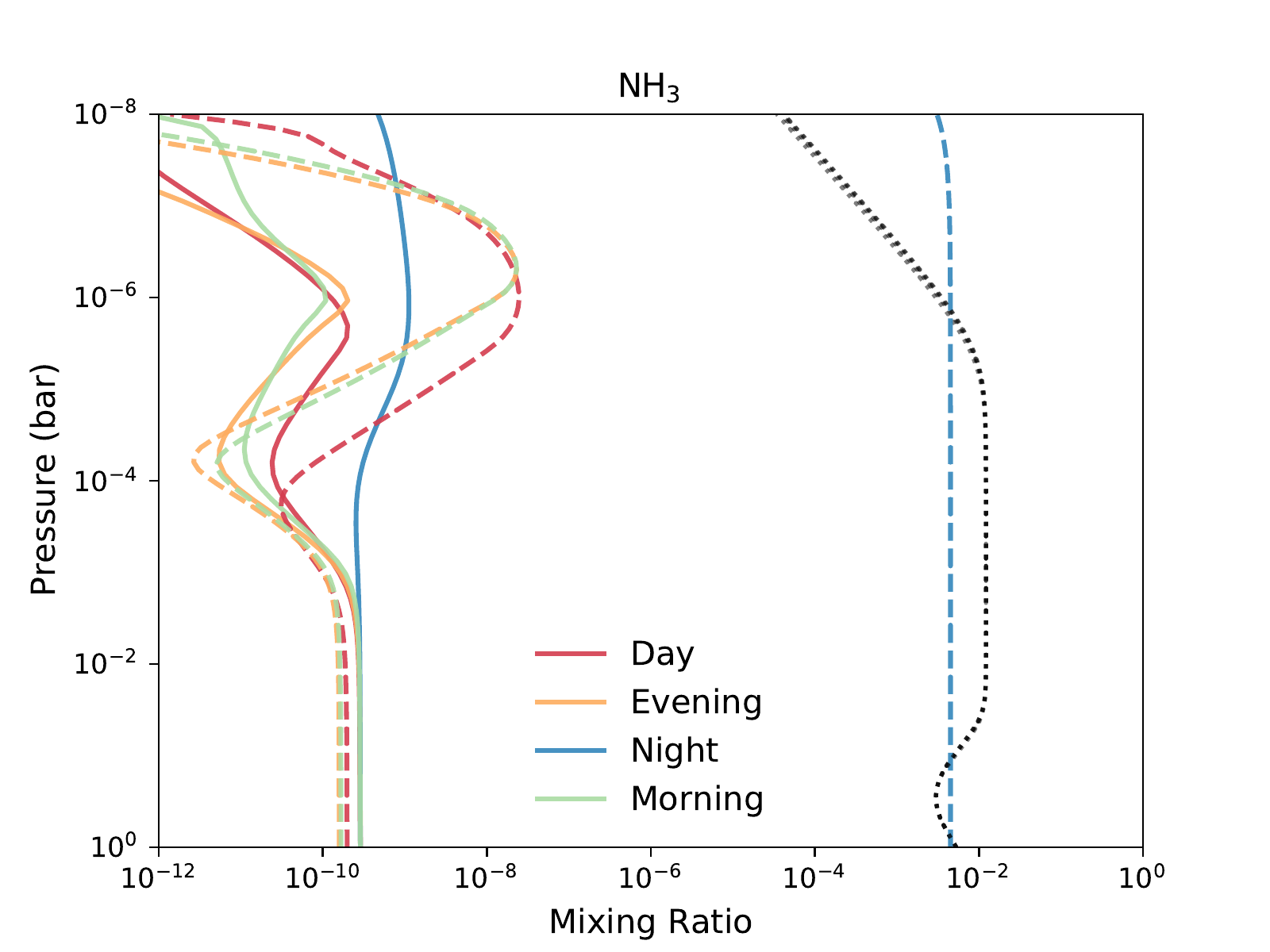}{\columnwidth}{(a)}\fig{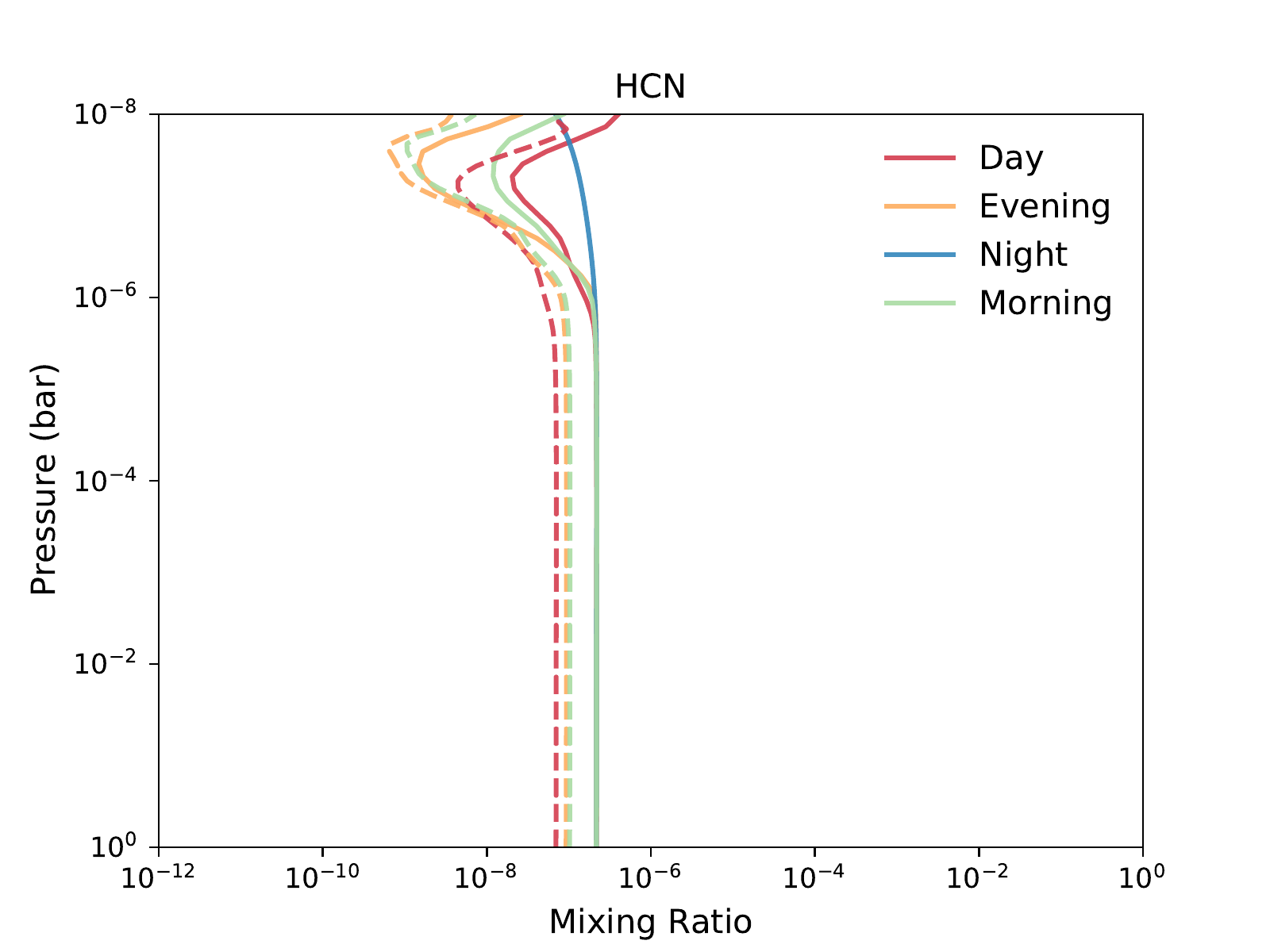}{\columnwidth}{(b)}}
\gridline{\fig{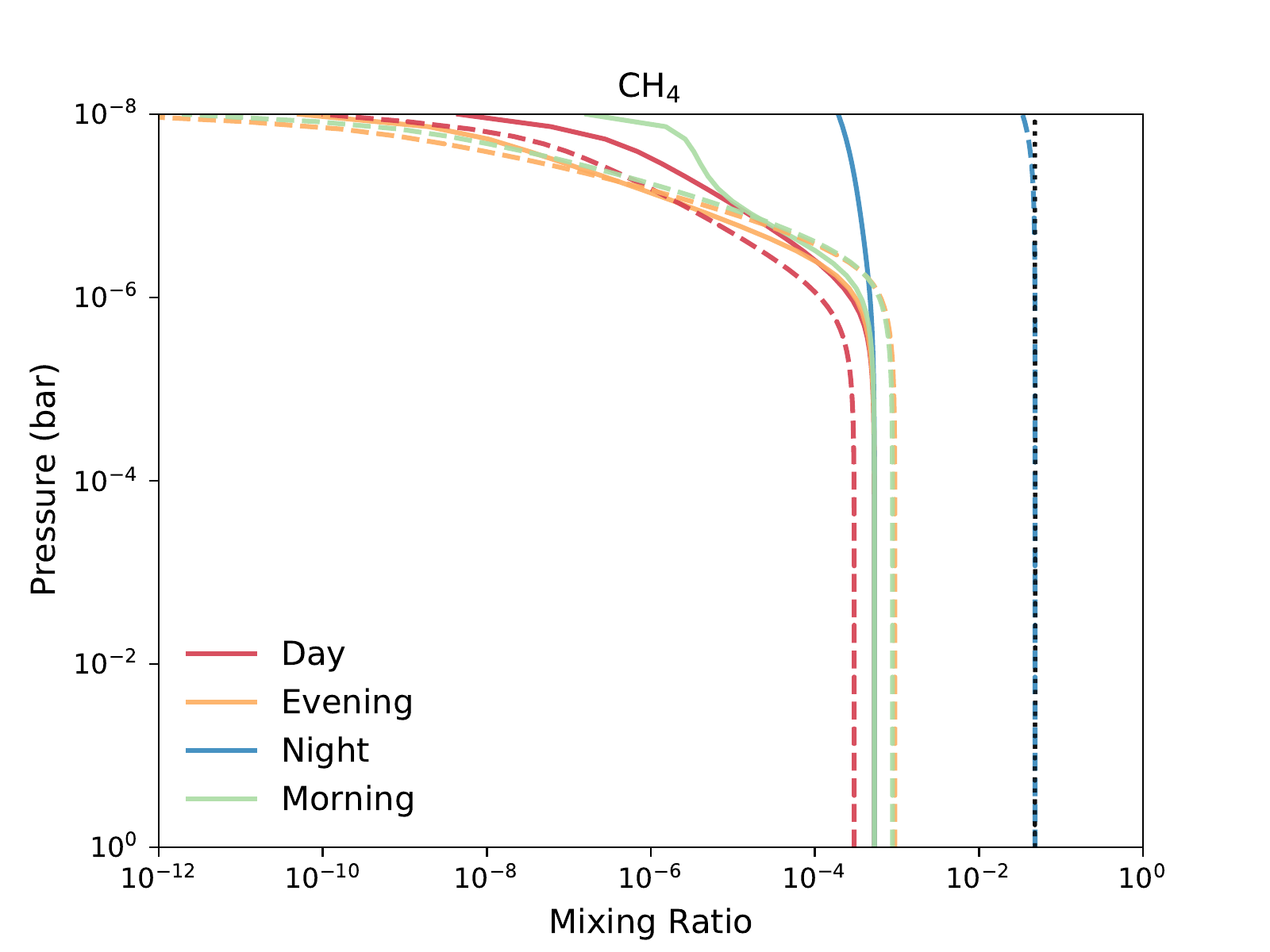}{\columnwidth}{(c)}
\fig{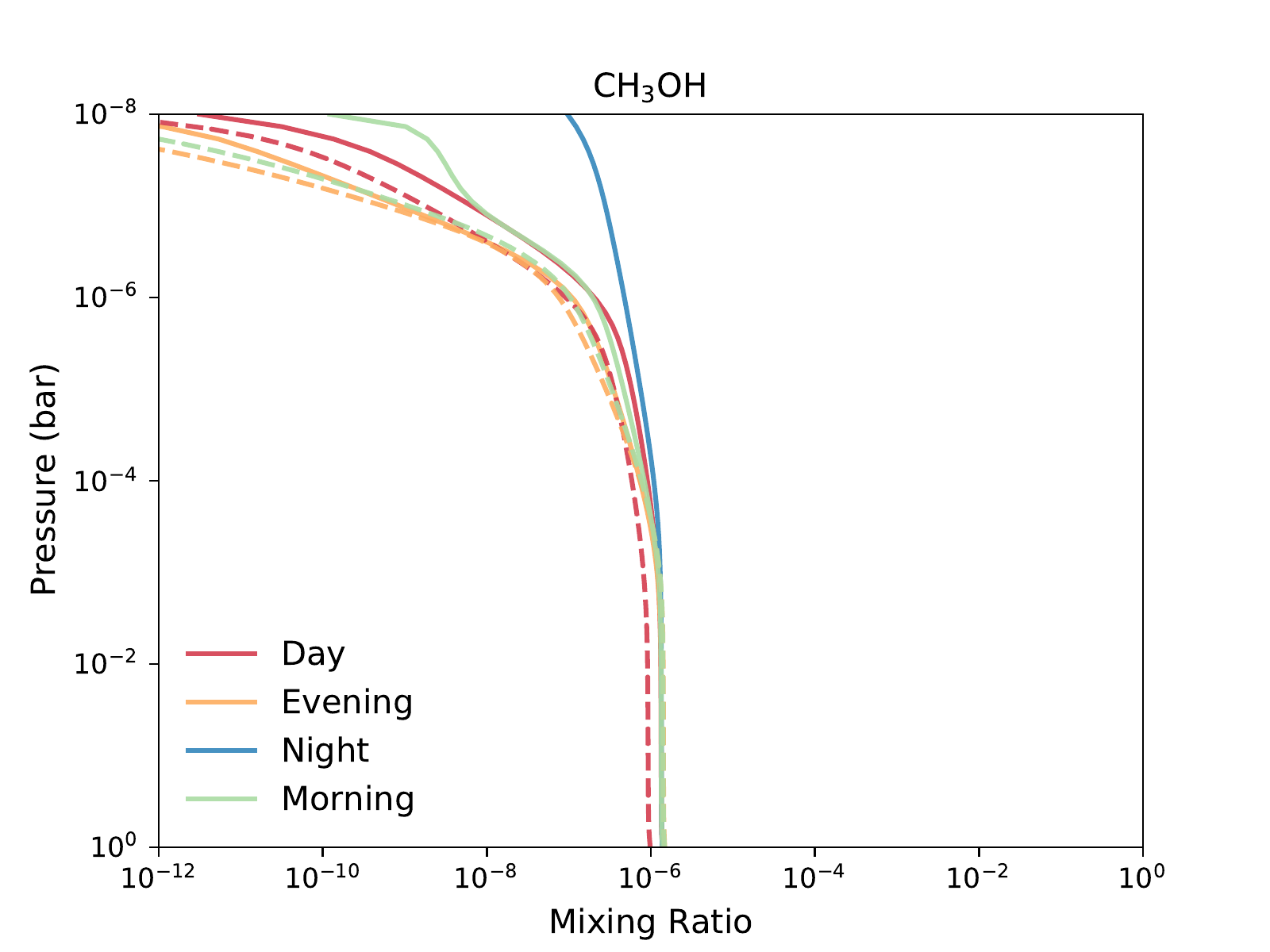}{\columnwidth}{(d)}}
\gridline{\fig{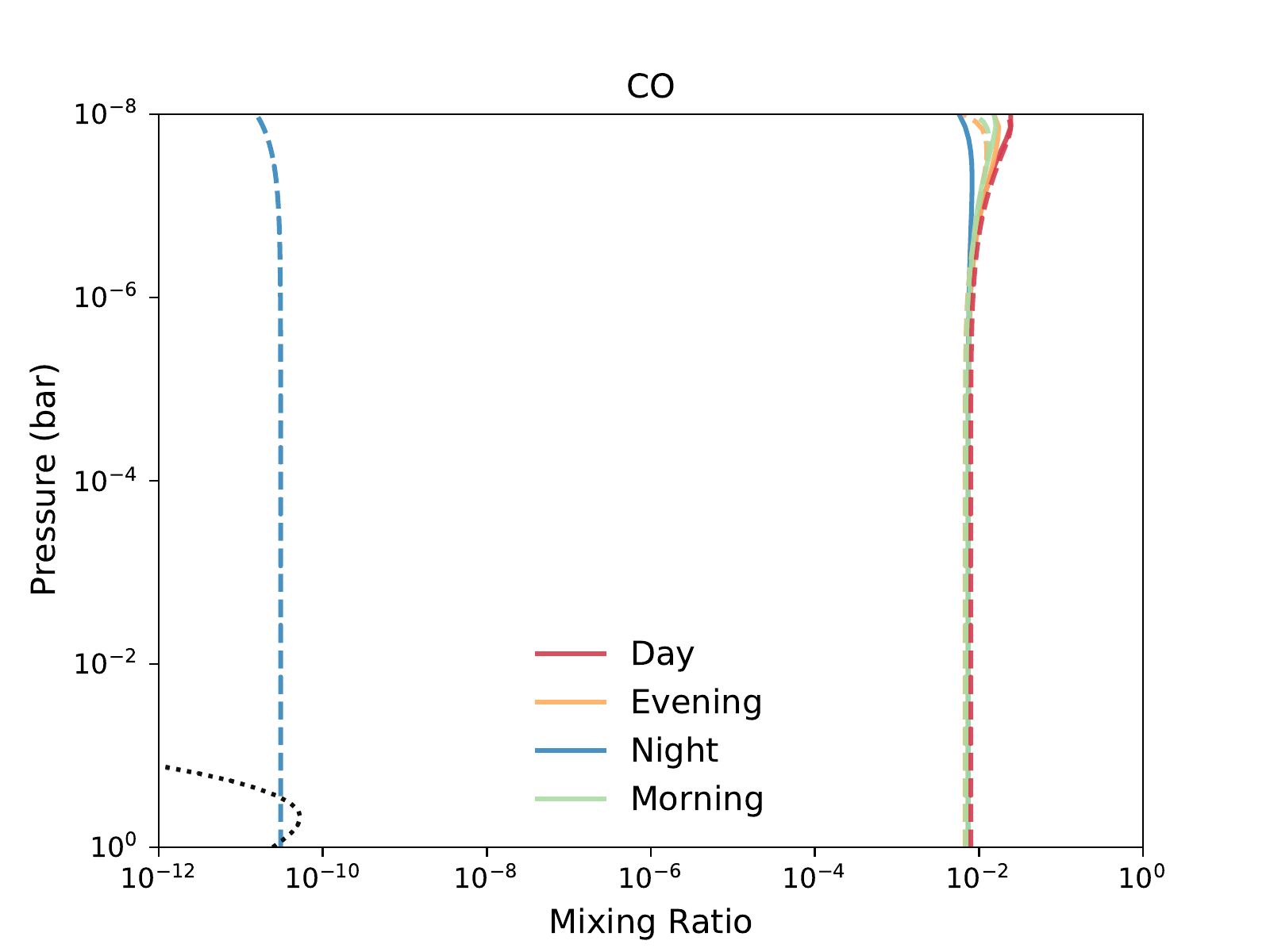}{\columnwidth}{(e)}
\fig{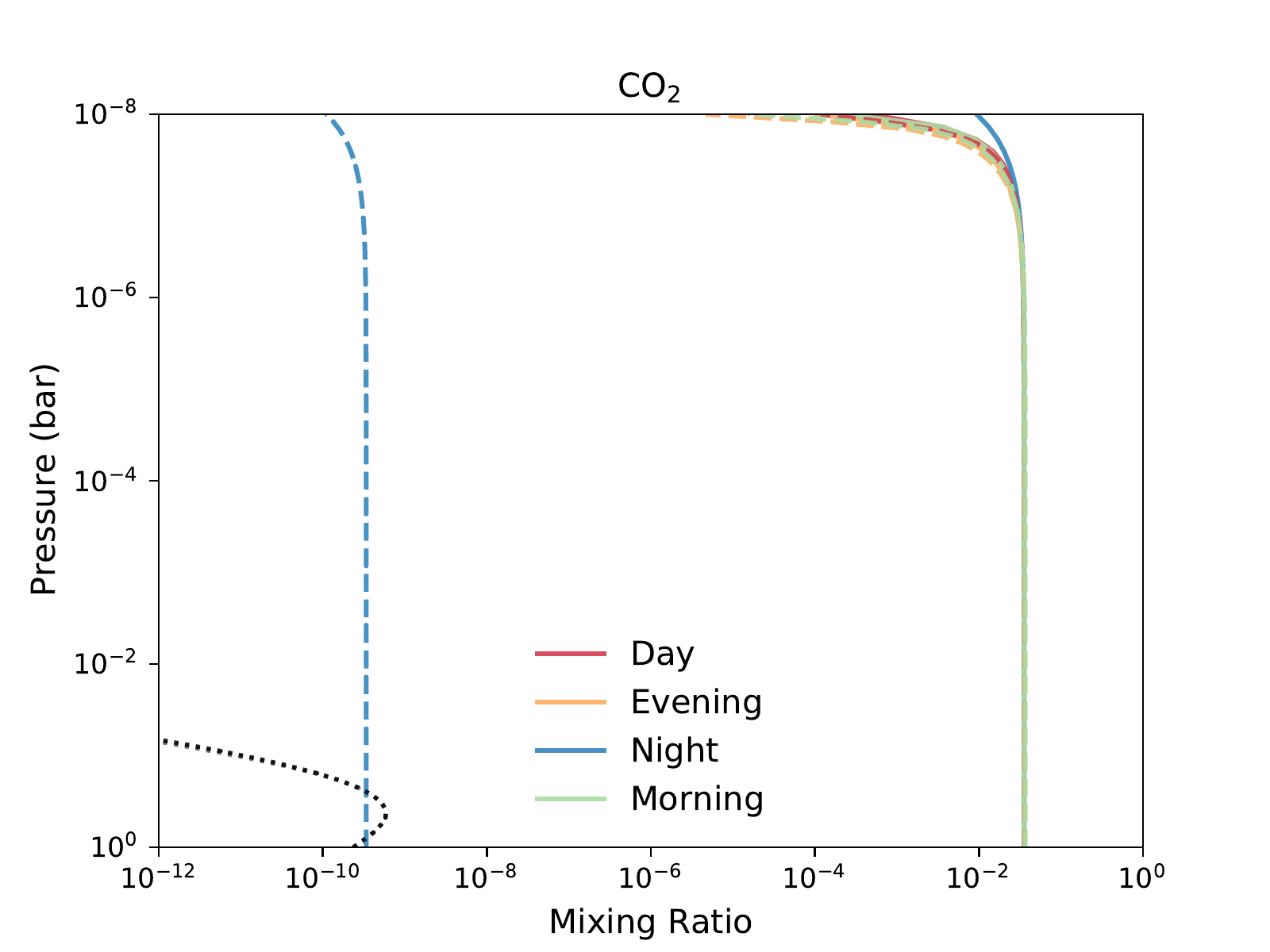}{\columnwidth}{(f)} }
\end{center}
\caption{The final mixing ratios of several species that can be potential surface indicators in the four quarters for an active M star, comparing simulations with (solid) and without (dashed) zonal wind. The black dotted lines indicate the equilibrium composition, which follows the global temperature.} 
\label{fig:2D}
\end{figure*}

We have established the long, $\mathcal{O}$(Myr) timescale of the nitrogen conversion in Section \ref{sec:1D_results}. On a tidally locked exoplanet, the conversion driven by photochemistry only occurs on the dayside but completely shuts down on the permanent nightside. Without photolysis, \ce{NH3} is able to maintain a high abundance on the nightside. To answer the question how the chemical evolution with a shallow surface would be regulated by the global circulation between the dayside and nightside, we apply a 2D photochemical model on a meridionally averaged equatorial plane to address the effects of day-night transport.

The temperature and wind for the 2D photochemical model are adopted from the 3D GCM output of K2-18b, whose global structure can be found in Figure \ref{fig:GCM}. We employ four quarters (Figure \ref{fig:GCM} (d)) in our 2D model for clarity, while the full 2D chemical output is included in the supplementary figure. To isolate the effects of horizontal transport, Figure \ref{fig:2D} compares the steady-state abundances with a 1-bar surface and an active M star to those without including the zonal wind. 

It is evident that without a recycling mechanism, the horizontal transport is able to exhaust \ce{NH3} on the nightside. Even the modest zonal wind ($\sim$ 1 m/s) yields a horizontal transport timescale less than a year, orders of magnitude faster than the chemical evolution. As a result, most species are homogenized below 1 mbar and exhibit abundances close to those predicted by the 1D model without day-night transport. The model for a quiet M star shows qualitatively consistent results (Figure \ref{fig:2D-5gyr}), with reduced photochemical destruction of \ce{NH3} and higher abundance of HCN. The results we find for the fiducial example of K2-18b can be generalized to most temperate, tidally-locked planets with a chemically inactive surface, since the equatorial jet is a robust dynamical property \citep[e.g.][]{Ray2019}.

\begin{figure}
\begin{center}
\includegraphics[width=0.94\columnwidth]{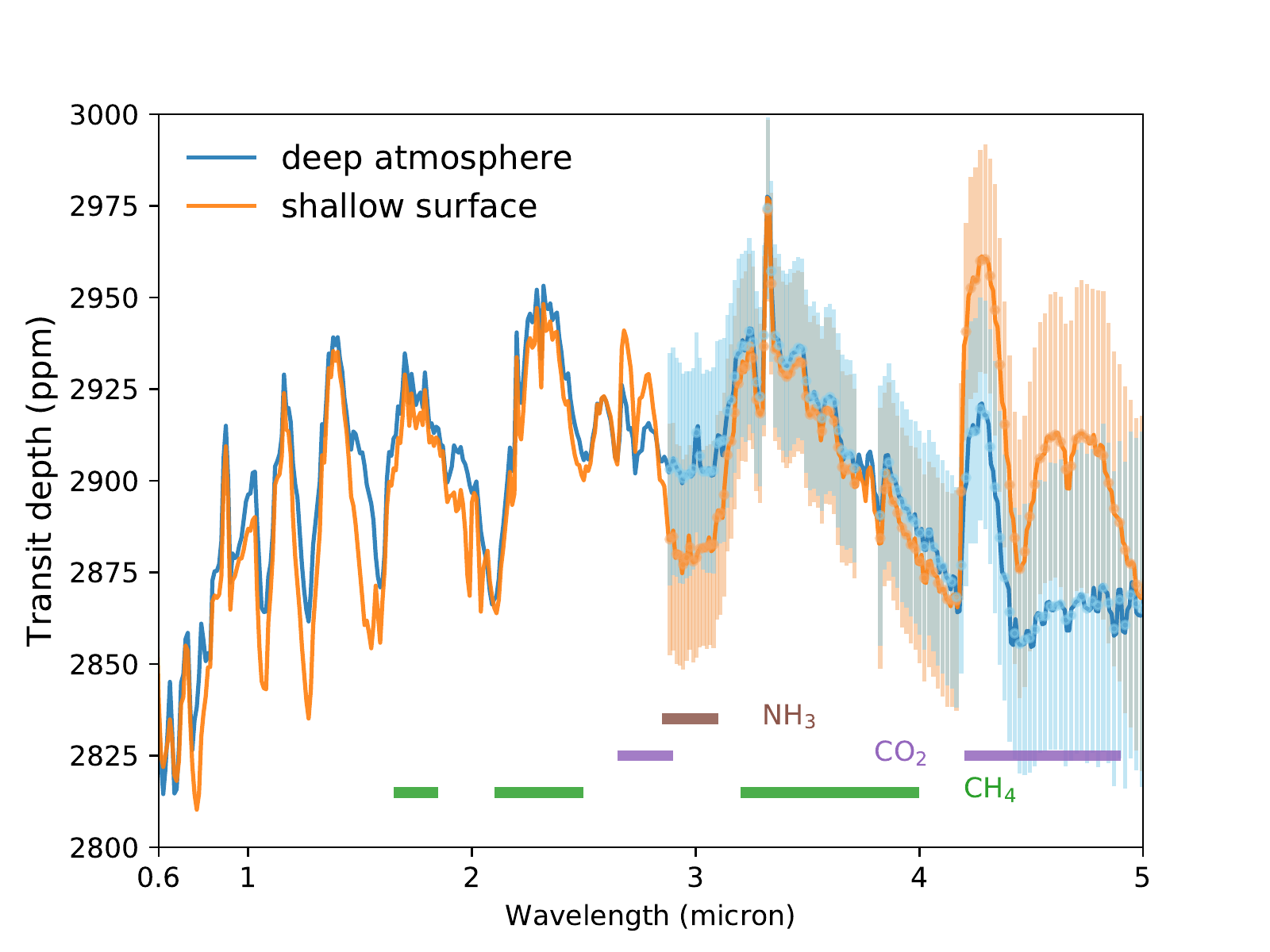}
\includegraphics[width=0.94\columnwidth]{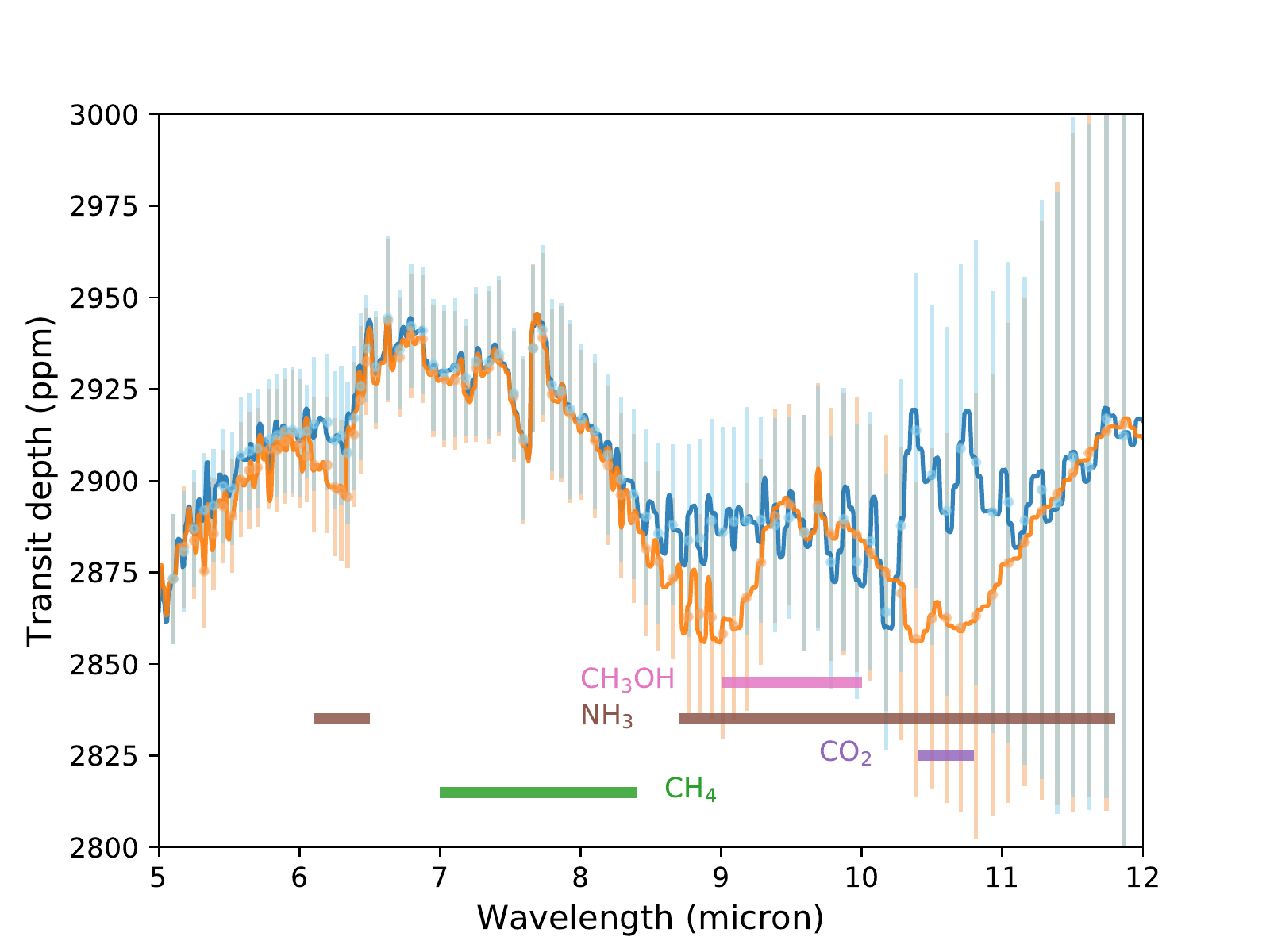}
\includegraphics[width=0.94\columnwidth]{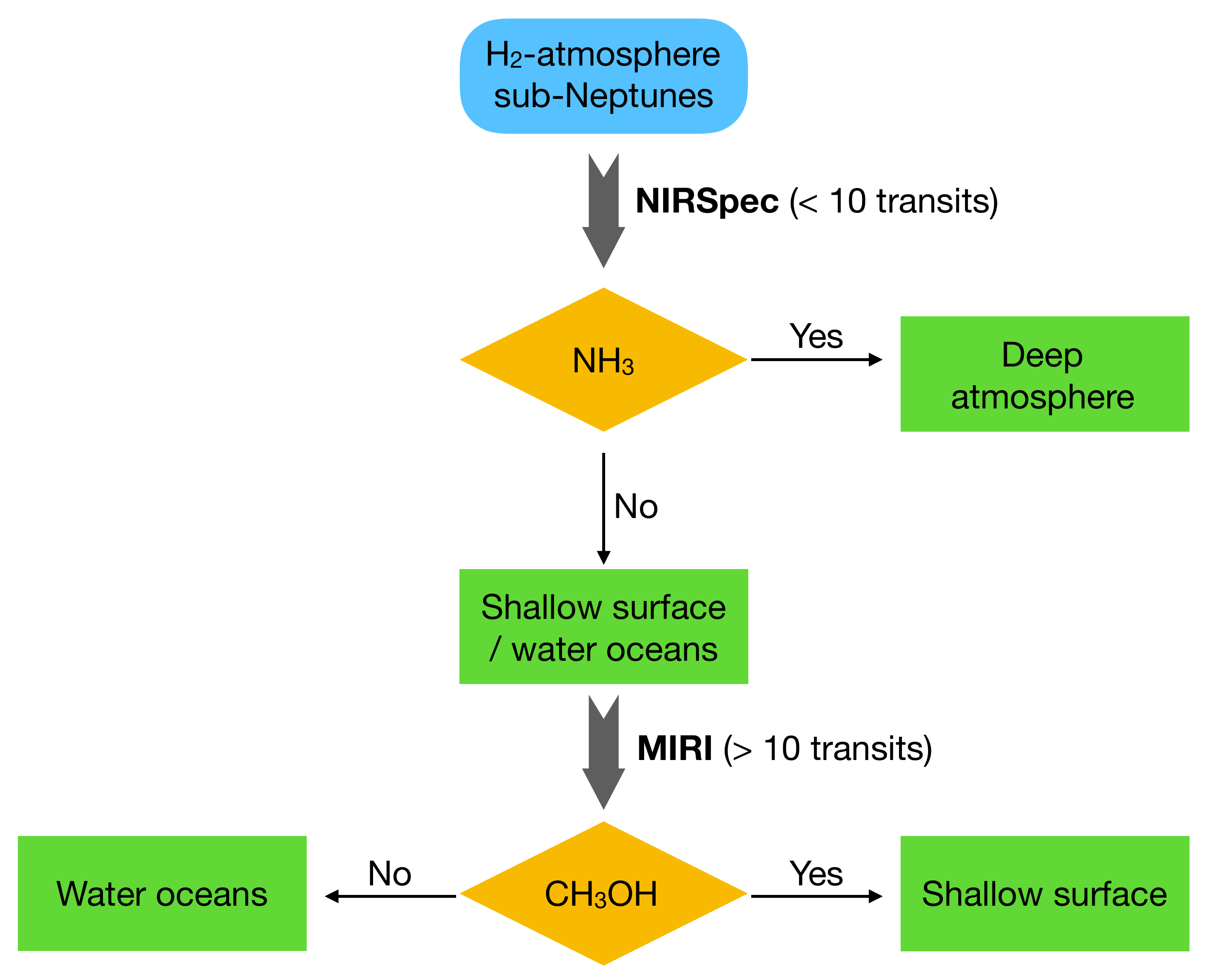}
\end{center}
\caption{The synthetic transmission spectra of K2-18b (top and middle panels) for the deep (1000-bar boundary; shown in blue) and shallow (1-bar surface; shown in orange) model. Points with error bars (light blue: deep atmosphere; light orange: 1-bar-surface) are simulated observations by the JWST NIRSpec G395H for 3 transits (top) and MIRI LRS for 20 transits (middle) without scatter produced by Pandexo \citep{Pandexo2017}. The absorption features of several molecules are indicated by the color bands. The bottom panel shows the flowchart to progressively identify shallow surfaces and ocean worlds with JWST instruments.}
\label{fig:spectra}
\end{figure}

\section{Spectral Indicator}
Since the 1D model for the dayside is verified to represent the globally uniform distribution, we compare the transmission spectra from our deep-atmosphere and shallow-surface models in Figure \ref{fig:spectra}. The most distinctive differences are \ce{CO2} absorption features at 4.2 -- 5 $\mu m$ and the lack of \ce{NH3} at 3, 8.8 -- 12 $\mu m$ for the shallow-surface case. Interestingly, \ce{CH3OH} absorbs between 9 and 10 $\mu m$, well within the \ce{NH3} band. To quantify the ability of detecting these molecules with JWST, we used Pandexo \citep{Pandexo2017} to generate the required observations to disentangle between our two models. We find that with it is possible to identify the \ce{NH3} feature at 3 $\mu m$ with only 3 transit observations using NIRSpec. When observing with MIRI,  it is possible to further discriminate \ce{CH3OH} from \ce{NH3} with around 20 transits. Conversely, HCN is not abundant enough in the region sensitive to transmission to show observable features.   

\section{Discussion}\label{sec:discussion}
Interactions with the planetary interior are not included in this work as we assume zero flux at the lower (surface) boundary for all species. Regarding \ce{NH3}, agricultural ammonia on Earth is absorbed by surfaces with a fairly short residence time  \citep[$\sim$ days, e.g.,][]{Seinfeld2016,Jia2016}. We have tested our model with Earth surface conditions assuming biotic processes are involved, and even the slowest dry deposition of \ce{NH3} on a desert \citep[$v_{dep} = 0.0002$ cm/s,][]{Jia2016} acts as a more efficient sink than the photochemical sink (Section \ref{sec:1D_results}). However, without the biosphere transferring ammonia to different organic nitrogen, it is conceivable that ammonia returns to the atmosphere at once and makes effectively net zero deposition. In general, various geological processes can be crucial in controlling trace gases in the atmosphere, especially determining the redox state of the atmosphere. For example, \cite{Wogan2020} find volcanic outgassing unlikely to be reducing, while \citet{2021ApJ...914L...4L} demonstrate that the interior of sub-Neptunes can remain reducing due to vigorous internal convection and \cite{Zahnle2020} suggest iron-rich impacts can also generate reducing atmospheres that favor \ce{NH3} and \ce{CH4}. In the case of an atmosphere with scattering clouds/hazes or less irradiated than K2-18b \citep{Piette2020,Blain2021}, the surface temperature can be brought down to suppress water evaporation \citep{Scheucher2020} and allow liquid water oceans, which can participate in regulating the inventory of soluble gases (e.g., \ce{NH3}, \ce{CO2}, and \ce{CH3OH}). With mobile lid tectonics \citep{2013Icar..225...50T,Meier2021}, the long-term evolution of \ce{CO2} in the shallow atmosphere case is expected to be governed by outgassing and weathering \citep[e.g.][]{Ray2010}.

Since \ce{CO2} generally has other geological sources and HCN is unsuitable as a surface proxy for its strong dependence on the stellar type and vertical mixing, we propose including \ce{CH3OH} as a complementary indicator along with \ce{NH3}. On Earth, \ce{CH3OH} is mainly produced by plants and deposited to the ocean \citep{Yang2013}, whereas in a \ce{H2}-dominated atmosphere with a shallow surface, \ce{CH3OH} is produced during the process of \ce{CH4} oxidation. Detection of \ce{NH3} without \ce{CH3OH} is consistent with the deep/no-surface scenario, while detection of \ce{CH3OH} but without \ce{NH3} indicates the presence of a surface. Non-detection of both \ce{NH3} and \ce{CH3OH} can imply a global water-ocean world \citep{Hu2021} as they are highly soluble in water. We summarize the flowchart of inferring the surface property of sub-Neptunes with \ce{H2}-dominated atmospheres using JWST in the bottom panel of Figure \ref{fig:spectra}.

\section{Summary}
1D modeling of sub-Neptune exoplanets suggests that \ce{NH3} is depleted below detectable level on the dayside in the presence of a shallow surface, while HCN can either increase or decrease depending on vertical mixing and quiet/active M stars. \ce{CH3OH} is found to be consistently enhanced for all shallow-surface simulations. We construct a 2D photochemical framework to account for the day-night circulation and find the global abundance is overall quenched from the dayside as the chemical conversion takes $\mathcal{O}$(Myr). Our results suggest that HCN is not applicable to determine the pressure level of the surface. Instead, the shallow-surface scenario can be ruled out by \ce{NH3} detection, whereas positive detection of \ce{CH3OH} with negative detection of \ce{NH3} indicate shallow and dry surfaces. 








\software{Numpy \citep{numpy}; Scipy \citep{scipy}; Matplotlib \citep{matplotlib}; VULCAN (\url{https://github.com/exoclime/VULCAN/releases/tag/v2.0}).}

\section*{acknowledgments}
S.-M.T. acknowledges support from the European community through the ERC advanced grant EXOCONDENSE (\#740963; PI: R.T. Pierrehumbert). T.L. has been supported by the Simons Foundation (SCOL award \#611576). This work has made use of the synthetic spectra from the HAZMAT program; \url{doi:10.17909/t9-j6bz-5g89}. We thank J. Moses for thoughtful discussion. 

%



\bibliography{master_bib}{} 

\begin{thebibliography}{}
\expandafter\ifx\csname natexlab\endcsname\relax\def\natexlab#1{#1}\fi
\providecommand{\url}[1]{\href{#1}{#1}}
\providecommand{\dodoi}[1]{doi:~\href{http://doi.org/#1}{\nolinkurl{#1}}}
\providecommand{\doeprint}[1]{\href{http://ascl.net/#1}{\nolinkurl{http://ascl.net/#1}}}
\providecommand{\doarXiv}[1]{\href{https://arxiv.org/abs/#1}{\nolinkurl{https://arxiv.org/abs/#1}}}

\bibitem[{{Aguichine} {et~al.}(2021){Aguichine}, {Mousis}, {Deleuil}, \&
  {Marcq}}]{Aguichine2021}
{Aguichine}, A., {Mousis}, O., {Deleuil}, M., \& {Marcq}, E. 2021, \apj, 914,
  84, \dodoi{10.3847/1538-4357/abfa99}

\bibitem[{{Ag{\'u}ndez} {et~al.}(2014){Ag{\'u}ndez}, {Venot}, {Selsis}, \&
  {Iro}}]{Agundez2014}
{Ag{\'u}ndez}, M., {Venot}, O., {Selsis}, F., \& {Iro}, N. 2014, \apj, 781, 68,
  \dodoi{10.1088/0004-637X/781/2/68}

\bibitem[{{Batalha} {et~al.}(2017){Batalha}, {Mandell}, {Pontoppidan},
  {Stevenson}, {Lewis}, {Kalirai}, {Earl}, {Greene}, {Albert}, \&
  {Nielsen}}]{Pandexo2017}
{Batalha}, N.~E., {Mandell}, A., {Pontoppidan}, K., {et~al.} 2017, \pasp, 129,
  064501, \dodoi{10.1088/1538-3873/aa65b0}

\bibitem[{{Bean} {et~al.}(2021){Bean}, {Raymond}, \& {Owen}}]{Bean2021}
{Bean}, J.~L., {Raymond}, S.~N., \& {Owen}, J.~E. 2021, Journal of Geophysical
  Research (Planets), 126, e06639, \dodoi{10.1029/2020JE006639}

\bibitem[{{Benneke} {et~al.}(2019){Benneke}, {Wong}, {Piaulet}, {Knutson},
  {Lothringer}, {Morley}, {Crossfield}, {Gao}, {Greene}, {Dressing},
  {Dragomir}, {Howard}, {McCullough}, {Kempton}, {Fortney}, \&
  {Fraine}}]{Benneke2019}
{Benneke}, B., {Wong}, I., {Piaulet}, C., {et~al.} 2019, \apjl, 887, L14,
  \dodoi{10.3847/2041-8213/ab59dc}

\bibitem[{{Blain, D.} {et~al.}(2021){Blain, D.}, {Charnay, B.}, \& {B\'ezard,
  B.}}]{Blain2021}
{Blain, D.}, {Charnay, B.}, \& {B\'ezard, B.} 2021, A\&A, 646, A15,
  \dodoi{10.1051/0004-6361/202039072}

\bibitem[{{Charnay} {et~al.}(2015){Charnay}, {Meadows}, \&
  {Leconte}}]{Charnay2015}
{Charnay}, B., {Meadows}, V., \& {Leconte}, J. 2015, \apj, 813, 15,
  \dodoi{10.1088/0004-637X/813/1/15}

\bibitem[{{Chubb} {et~al.}(2021){Chubb}, {Rocchetto}, {Yurchenko}, {Min},
  {Waldmann}, {Barstow}, {Molli{\`e}re}, {Al-Refaie}, {Phillips}, \&
  {Tennyson}}]{2021Chubb}
{Chubb}, K.~L., {Rocchetto}, M., {Yurchenko}, S.~N., {et~al.} 2021, \aap, 646,
  A21, \dodoi{10.1051/0004-6361/202038350}

\bibitem[{Coles {et~al.}(2019)Coles, , Yurchenko, \& Tennyson}]{NH3_opacity}
Coles, P.~A., , Yurchenko, S.~N., \& Tennyson, J. 2019, Mon. Not. R. Astron.
  Soc., 490, 4638, \dodoi{10.1093/mnras/stz2778}

\bibitem[{{Drummond} {et~al.}(2020){Drummond}, {H{\'e}brard}, {Mayne}, {Venot},
  {Ridgway}, {Changeat}, {Tsai}, {Manners}, {Tremblin}, {Abraham}, {Sing}, \&
  {Kohary}}]{Drummond2020}
{Drummond}, B., {H{\'e}brard}, E., {Mayne}, N.~J., {et~al.} 2020, \aap, 636,
  A68, \dodoi{10.1051/0004-6361/201937153}

\bibitem[{Feng {et~al.}(2020)Feng, Line, \& Fortney}]{Feng2020}
Feng, Y.~K., Line, M.~R., \& Fortney, J.~J. 2020, The Astronomical Journal,
  160, 137, \dodoi{10.3847/1538-3881/aba8f9}

\bibitem[{{Fortney} {et~al.}(2013){Fortney}, {Mordasini}, {Nettelmann},
  {Kempton}, {Greene}, \& {Zahnle}}]{Fortney2013}
{Fortney}, J.~J., {Mordasini}, C., {Nettelmann}, N., {et~al.} 2013, \apj, 775,
  80, \dodoi{10.1088/0004-637X/775/1/80}

\bibitem[{Gordon \& {et al.}(2017)}]{CH3OH_opacity_1}
Gordon, I.~E., \& {et al.} 2017, J. Quant. Spectrosc. Radiat. Transf., 203, 3,
  \dodoi{10.1016/j.jqsrt.2017.06.038}

\bibitem[{Hammond \& {Pierrehumbert}(2017)}]{Hammond2017}
Hammond, M., \& {Pierrehumbert}, R.~T. 2017, The Astrophysical Journal, 849,
  152, \dodoi{10.3847/1538-4357/aa9328}

\bibitem[{Hammond \& Pierrehumbert(2018)}]{Hammond2018}
Hammond, M., \& Pierrehumbert, R.~T. 2018, The Astrophysical Journal, 869, 65,
  \dodoi{10.3847/1538-4357/aaec03}

\bibitem[{{Hsu} {et~al.}(2019){Hsu}, {Ford}, {Ragozzine}, \& {Ashby}}]{Hsu2019}
{Hsu}, D.~C., {Ford}, E.~B., {Ragozzine}, D., \& {Ashby}, K. 2019, \aj, 158,
  109, \dodoi{10.3847/1538-3881/ab31ab}

\bibitem[{Hu {et~al.}(2021)Hu, Damiano, Scheucher, Kite, Seager, \&
  Rauer}]{Hu2021}
Hu, R., Damiano, M., Scheucher, M., {et~al.} 2021, 921, L8,
  \dodoi{10.3847/2041-8213/ac1f92}

\bibitem[{Hunter(2007)}]{matplotlib}
Hunter, J.~D. 2007, Computing in Science Engineering, 9, 90,
  \dodoi{10.1109/MCSE.2007.55}

\bibitem[{{Irwin} {et~al.}(2008){Irwin}, {Teanby}, {de Kok}, {Fletcher},
  {Howett}, {Tsang}, {Wilson}, {Calcutt}, {Nixon}, \& {Parrish}}]{Irwin2008}
{Irwin}, P.~G.~J., {Teanby}, N.~A., {de Kok}, R., {et~al.} 2008, \jqsrt, 109,
  1136, \dodoi{10.1016/j.jqsrt.2007.11.006}

\bibitem[{{Jia} {et~al.}(2016){Jia}, {Yu}, {Gao}, {He}, {Wang}, {Jiao}, \&
  {Zuo}}]{Jia2016}
{Jia}, Y., {Yu}, G., {Gao}, Y., {et~al.} 2016, Scientific Reports, 6, 19810,
  \dodoi{10.1038/srep19810}

\bibitem[{Lee {et~al.}(2021)Lee, Parmentier, Hammond, Grimm, Kitzmann, Tan,
  Tsai, \& Pierrehumbert}]{Lee2021}
Lee, E. K.~H., Parmentier, V., Hammond, M., {et~al.} 2021, MNRAS, 000, 1.
\newblock \doarXiv{2106.11664}

\bibitem[{Lee {et~al.}(2020)Lee, Casewell, Chubb, Hammond, Tan, Tsai, \&
  Pierrehumbert}]{Lee2020}
Lee, G.~K., Casewell, S.~L., Chubb, K.~L., {et~al.} 2020, Monthly Notices of
  the Royal Astronomical Society, 496, 4674, \dodoi{10.1093/mnras/staa1882}

\bibitem[{Li {et~al.}({2015})Li, Gordon, Rothman, Tan, Hu, Kassi, Campargue, \&
  Medvedev}]{CO_opacity}
Li, G., Gordon, I.~E., Rothman, L.~S., {et~al.} {2015}, Astrophys. J. Suppl.,
  {216}, 15, \dodoi{10.1088/0067-0049/216/1/15}

\bibitem[{{Lichtenberg}(2021)}]{2021ApJ...914L...4L}
{Lichtenberg}, T. 2021, \apjl, 914, L4, \dodoi{10.3847/2041-8213/ac0146}

\bibitem[{Lindzen(1981)}]{Lindzen1981}
Lindzen, R.~S. 1981, Journal of Geophysical Research: Oceans, 86, 9707,
  \dodoi{https://doi.org/10.1029/JC086iC10p09707}

\bibitem[{Lopez \& Fortney(2013)}]{Lopez2013}
Lopez, E.~D., \& Fortney, J.~J. 2013, The Astrophysical Journal, 776, 2,
  \dodoi{10.1088/0004-637x/776/1/2}

\bibitem[{{Madhusudhan} {et~al.}(2020){Madhusudhan}, {Nixon}, {Welbanks},
  {Piette}, \& {Booth}}]{Madhu2020}
{Madhusudhan}, N., {Nixon}, M.~C., {Welbanks}, L., {Piette}, A. A.~A., \&
  {Booth}, R.~A. 2020, \apjl, 891, L7, \dodoi{10.3847/2041-8213/ab7229}

\bibitem[{{Malik} {et~al.}(2019b){Malik}, {Kempton}, {Koll}, {Mansfield},
  {Bean}, \& {Kite}}]{Malik2019b}
{Malik}, M., {Kempton}, E. M.~R., {Koll}, D. D.~B., {et~al.} 2019b, \apj, 886,
  142, \dodoi{10.3847/1538-4357/ab4a05}

\bibitem[{Malik {et~al.}(2019a)Malik, Kitzmann, Mendon{\c{c}}a, Grimm, Marleau,
  Linder, Tsai, \& Heng}]{Malik2019}
Malik, M., Kitzmann, D., Mendon{\c{c}}a, J.~M., {et~al.} 2019a, The
  Astronomical Journal, 157, 170, \dodoi{10.3847/1538-3881/ab1084}

\bibitem[{{May} \& {Rauscher}(2020)}]{May2020}
{May}, E.~M., \& {Rauscher}, E. 2020, \apj, 893, 161,
  \dodoi{10.3847/1538-4357/ab838b}

\bibitem[{{Meier} {et~al.}(2021){Meier}, {Bower}, {Lichtenberg}, {Tackley}, \&
  {Demory}}]{Meier2021}
{Meier}, T.~G., {Bower}, D.~J., {Lichtenberg}, T., {Tackley}, P.~J., \&
  {Demory}, B.-O. 2021, \apjl, 908, L48, \dodoi{10.3847/2041-8213/abe400}

\bibitem[{Moses {et~al.}(2013)Moses, Line, Visscher, Richardson, Nettelmann,
  Fortney, Barman, Stevenson, \& Madhusudhan}]{Moses2013}
Moses, J.~I., Line, M.~R., Visscher, C., {et~al.} 2013, The Astrophysical
  Journal, 777, 34, \dodoi{10.1088/0004-637x/777/1/34}

\bibitem[{Moses {et~al.}(2016)Moses, Marley, Zahnle, Line, Fortney, Barman,
  Visscher, Lewis, \& Wolff}]{Moses2016}
Moses, J.~I., Marley, M.~S., Zahnle, K., {et~al.} 2016, The Astrophysical
  Journal, 829, 66, \dodoi{10.3847/0004-637x/829/2/66}

\bibitem[{Oliphant(2007)}]{scipy}
Oliphant, T.~E. 2007, Computing in Science Engineering, 9, 10,
  \dodoi{10.1109/MCSE.2007.58}

\bibitem[{{Peacock} {et~al.}(2020){Peacock}, {Barman}, {Shkolnik}, {Loyd},
  {Schneider}, {Pagano}, \& {Meadows}}]{Peacock2020}
{Peacock}, S., {Barman}, T., {Shkolnik}, E.~L., {et~al.} 2020, \apj, 895, 5,
  \dodoi{10.3847/1538-4357/ab893a}

\bibitem[{{Pierrehumbert}(2010)}]{Ray2010}
{Pierrehumbert}, R.~T. 2010, {Principles of Planetary Climate}

\bibitem[{Pierrehumbert \& Ding(2016)}]{Pierrehumbert2016}
Pierrehumbert, R.~T., \& Ding, F. 2016, Proceedings of the Royal Society A:
  Mathematical, Physical and Engineering Sciences, 472,
  \dodoi{10.1098/rspa.2016.0107}

\bibitem[{Pierrehumbert \& Hammond(2019)}]{Ray2019}
Pierrehumbert, R.~T., \& Hammond, M. 2019, Annual Review of Fluid Mechanics,
  51, 275, \dodoi{10.1146/annurev-fluid-010518-040516}

\bibitem[{{Piette} \& {Madhusudhan}(2020)}]{Piette2020}
{Piette}, A. A.~A., \& {Madhusudhan}, N. 2020, \apj, 904, 154,
  \dodoi{10.3847/1538-4357/abbfb1}

\bibitem[{Scheucher {et~al.}(2020)Scheucher, Wunderlich, Grenfell, Godolt,
  Schreier, Kappel, Haus, Herbst, \& Rauer}]{Scheucher2020}
Scheucher, M., Wunderlich, F., Grenfell, J.~L., {et~al.} 2020, 898, 44,
  \dodoi{10.3847/1538-4357/ab9084}

\bibitem[{Seinfeld \& Pandis(2016)}]{Seinfeld2016}
Seinfeld, J.~H., \& Pandis, S.~N. 2016, Atmospheric chemistry and physics: from
  air pollution to climate change (John Wiley \&amp; Sons, Inc.)

\bibitem[{{Tackley} {et~al.}(2013){Tackley}, {Ammann}, {Brodholt}, {Dobson}, \&
  {Valencia}}]{2013Icar..225...50T}
{Tackley}, P.~J., {Ammann}, M., {Brodholt}, J.~P., {Dobson}, D.~P., \&
  {Valencia}, D. 2013, \icarus, 225, 50, \dodoi{10.1016/j.icarus.2013.03.013}

\bibitem[{Tsai {et~al.}(2018)Tsai, Kitzmann, Lyons, Mendon{\c{c}}a, Grimm, \&
  Heng}]{tsai18}
Tsai, S.-M., Kitzmann, D., Lyons, J.~R., {et~al.} 2018, ApJ, 862, 31,
  \dodoi{10.3847/1538-4357/aac834}

\bibitem[{Tsai {et~al.}(2017)Tsai, Lyons, Grosheintz, Rimmer, Kitzmann, \&
  Heng}]{tsai17}
Tsai, S.-M., Lyons, J.~R., Grosheintz, L., {et~al.} 2017, Astrophys. J. Suppl.
  Ser., 228, 1, \dodoi{10.3847/1538-4365/228/2/20}

\bibitem[{{Tsai} {et~al.}(2021){Tsai}, {Malik}, {Kitzmann}, {Lyons}, {Fateev},
  {Lee}, \& {Heng}}]{Tsai2021}
{Tsai}, S.-M., {Malik}, M., {Kitzmann}, D., {et~al.} 2021, arXiv e-prints,
  arXiv:2108.01790.
\newblock \doarXiv{2108.01790}

\bibitem[{{Tsiaras} {et~al.}(2019){Tsiaras}, {Waldmann}, {Tinetti}, {Tennyson},
  \& {Yurchenko}}]{Tsiaras2019}
{Tsiaras}, A., {Waldmann}, I.~P., {Tinetti}, G., {Tennyson}, J., \&
  {Yurchenko}, S.~N. 2019, Nature Astronomy, 3, 1086,
  \dodoi{10.1038/s41550-019-0878-9}

\bibitem[{van~der Walt {et~al.}(2011)van~der Walt, Colbert, \&
  Varoquaux}]{numpy}
van~der Walt, S., Colbert, S.~C., \& Varoquaux, G. 2011, Computing in Science
  Engineering, 13, 22, \dodoi{10.1109/MCSE.2011.37}

\bibitem[{{Wardenier} {et~al.}(2021){Wardenier}, {Parmentier}, {Lee}, {Line},
  \& {Gharib-Nezhad}}]{Joost2021}
{Wardenier}, J.~P., {Parmentier}, V., {Lee}, E. K.~H., {Line}, M.~R., \&
  {Gharib-Nezhad}, E. 2021, \mnras, 506, 1258, \dodoi{10.1093/mnras/stab1797}

\bibitem[{Wogan {et~al.}(2020)Wogan, Krissansen-Totton, \& Catling}]{Wogan2020}
Wogan, N., Krissansen-Totton, J., \& Catling, D.~C. 2020, The Planetary Science
  Journal, 1, 58, \dodoi{10.3847/psj/abb99e}

\bibitem[{{Yang} {et~al.}(2013){Yang}, {Nightingale}, {Beale}, {Liss},
  {Blomquist}, \& {Fairall}}]{Yang2013}
{Yang}, M., {Nightingale}, P.~D., {Beale}, R., {et~al.} 2013, Proceedings of
  the National Academy of Science, 110, 20034, \dodoi{10.1073/pnas.1317840110}

\bibitem[{{Yu} {et~al.}(2021){Yu}, {Moses}, {Fortney}, \& {Zhang}}]{Yu2021}
{Yu}, X., {Moses}, J.~I., {Fortney}, J.~J., \& {Zhang}, X. 2021, \apj, 914, 38,
  \dodoi{10.3847/1538-4357/abfdc7}

\bibitem[{Yurchenko {et~al.}(2020)Yurchenko, Mellor, Freedman, \&
  Tennyson}]{CO2_opacity}
Yurchenko, S.~N., Mellor, T.~M., Freedman, R.~S., \& Tennyson, J. 2020, Mon.
  Not. R. Astron. Soc., 496, 5282, \dodoi{10.1093/mnras/staa1874}

\bibitem[{Zahnle {et~al.}(2020)Zahnle, Lupu, Catling, \& Wogan}]{Zahnle2020}
Zahnle, K.~J., Lupu, R., Catling, D.~C., \& Wogan, N. 2020, The Planetary
  Science Journal, 1, 11, \dodoi{10.3847/psj/ab7e2c}

\bibitem[{{Zilinskas} {et~al.}(2020){Zilinskas}, {Miguel}, {Molli{\`e}re}, \&
  {Tsai}}]{Zilinskas2020}
{Zilinskas}, M., {Miguel}, Y., {Molli{\`e}re}, P., \& {Tsai}, S.-M. 2020,
  \mnras, 494, 1490, \dodoi{10.1093/mnras/staa724}

\end{thebibliography}
\bibliographystyle{aasjournal}


\pagebreak

\appendix




\section{Radiative feedback from disequilibrium chemistry}
Since the temperature structure is fixed by thermochemical equilibrium, we have also checked the radiative feedback as a result of disequilibrium chemistry. We re-run the 1D radiative-convective calculations with the final disequilibrium composition from Section \ref{sec:1D_results}. We find the resulting temperature from disequilibrium chemistry can be about 100 K lower at most in the convective zone above 1 bar, mainly from the decrease of opacity-predominating water. We have performed sensitivity tests and found the temperature difference does not alter the presented results (also see the thermal structure variance explored in \cite{Yu2021}). We will leave the effects of water condensation for future work.


\section{Running 2D photochemical models for millions years}
The horizontal transport flux by the zonal wind in Equation (\ref{eq:master}) yields a hyperbolic partial differential equation. The time step is ought to follow the same format as Courant–Friedrichs–Lewy (CFL) condition when numerically integrating the system. In other words, the time step must not exceed the time that zonal flow travels across adjacent vertical columns. For K2-18b, the time step limit is $\frac{\Delta x}{u} \sim 10^5$ s, which means more than $10^8$ integration steps are required to integrate the system to a million year for the long-term chemical evolution. To circumvent this computational load, we arbitrarily slow down the zonal wind (e.g., by 1000 times) such that a larger time step can be adopted. Once the chemistry has evolved after the Myr integration, we recover the correct wind velocity and run the model to final steady state. This seemingly risky approach can be justified by the timescale argument: The horizontal transport effectively interacts with vertical mixing and chemical evolution , which manifest drastically different timescales. The timescale of vertical mixing is within hours and the chemical evolution takes $\sim$ Myr, whereas the nominal horizontal transport has a timescale of $\sim$ days. Therefore, the {\it long-term compositional evolution} is expected to be qualitatively unaffected for any horizontal transport much slower than vertical mixing but orders of magnitude faster than the chemical evolution. That is, the tuned-down zonal wind still plays the role of passing on the chemical evolution from dayside to nightside. Finally, we have confirmed this approach by comparing the simulation at about 5000 years to that with 1000 times slower zonal wind and found no differences.

\begin{figure*}[ht!]
\begin{center}
\gridline{\fig{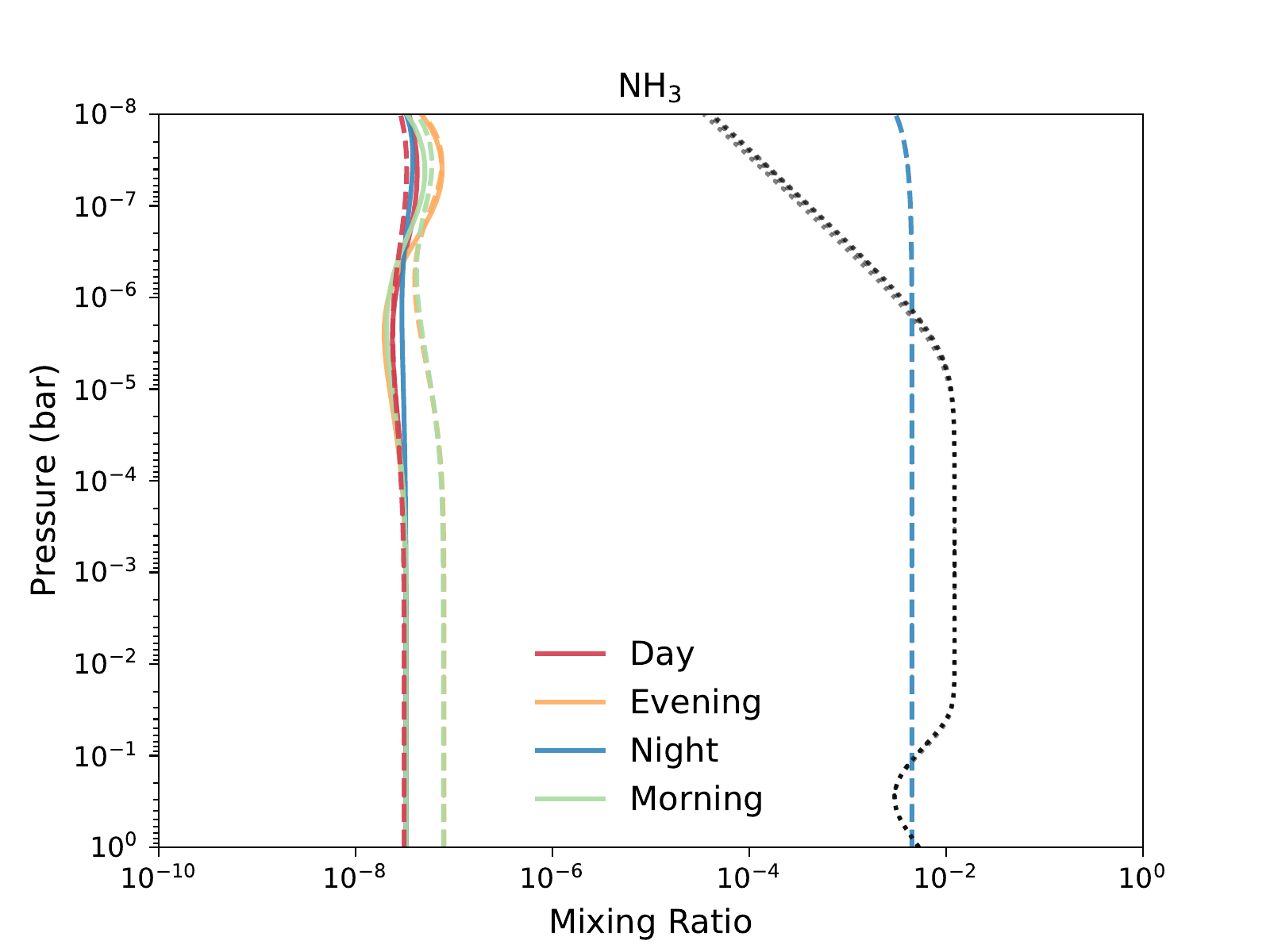}{0.5\columnwidth}{(a)} 
\fig{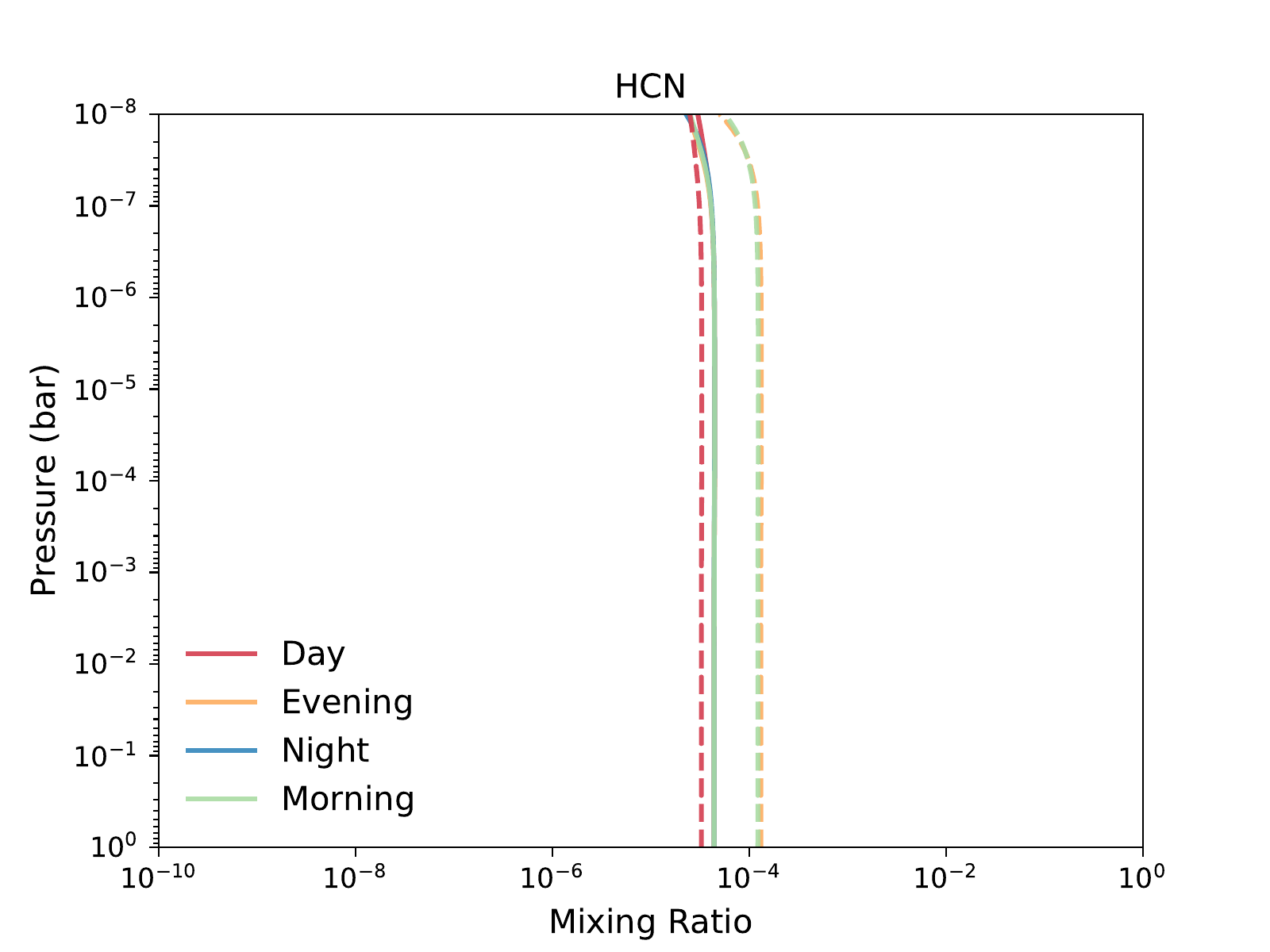}{0.5\columnwidth}{(b)}}
\gridline{\fig{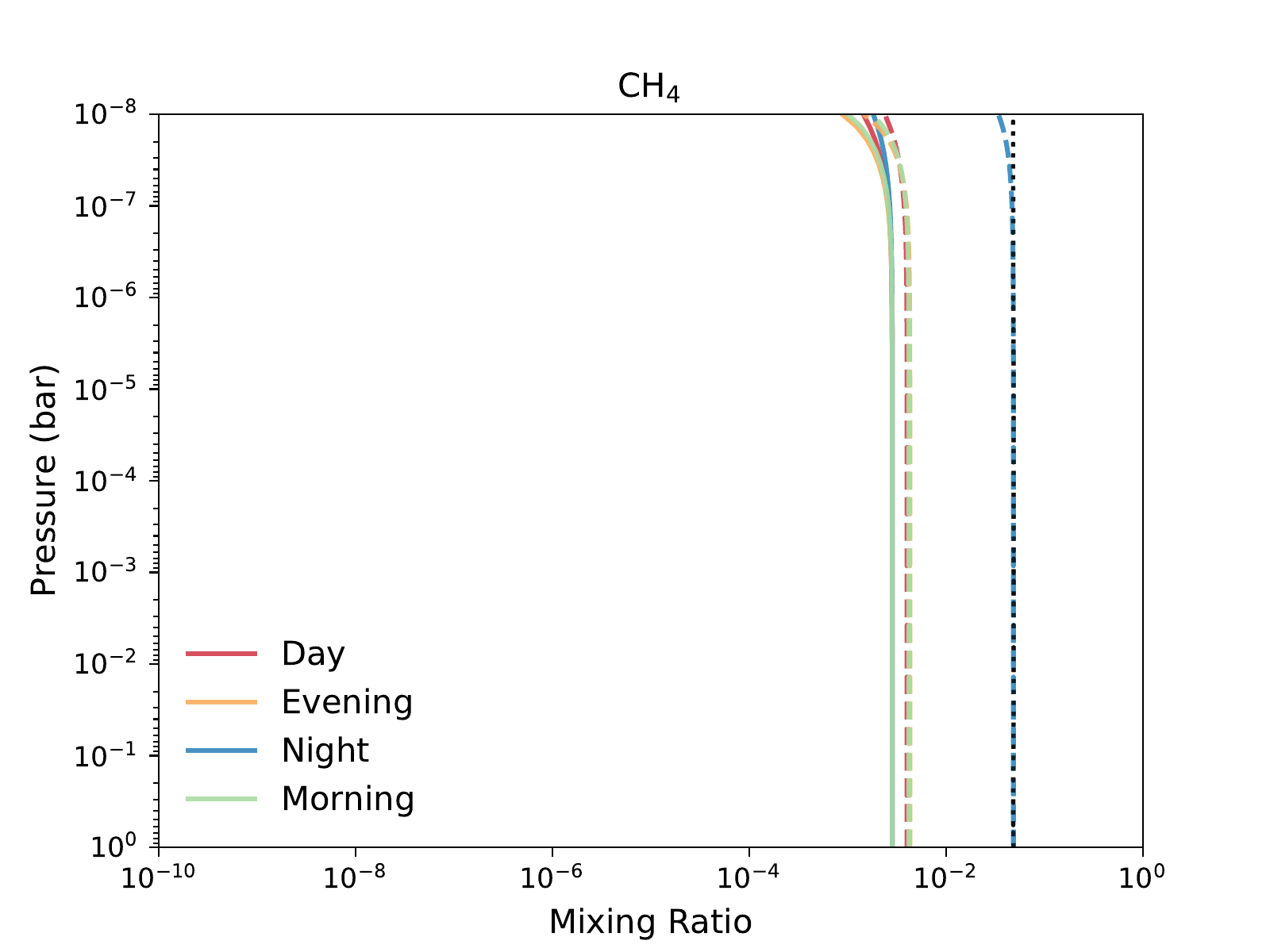}{0.5\columnwidth}{(c)}
\fig{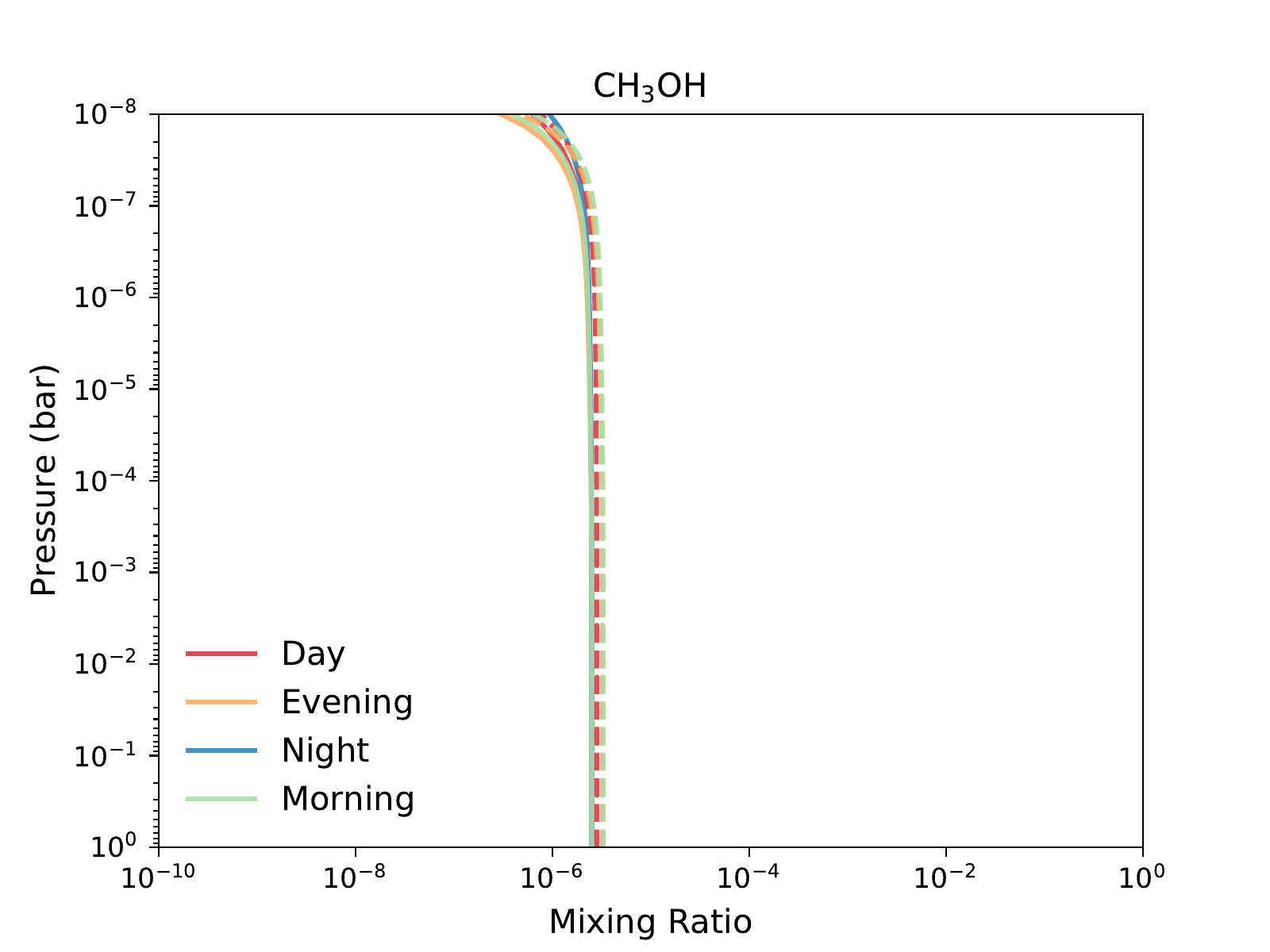}{0.5\columnwidth}{(d)}}
\gridline{\fig{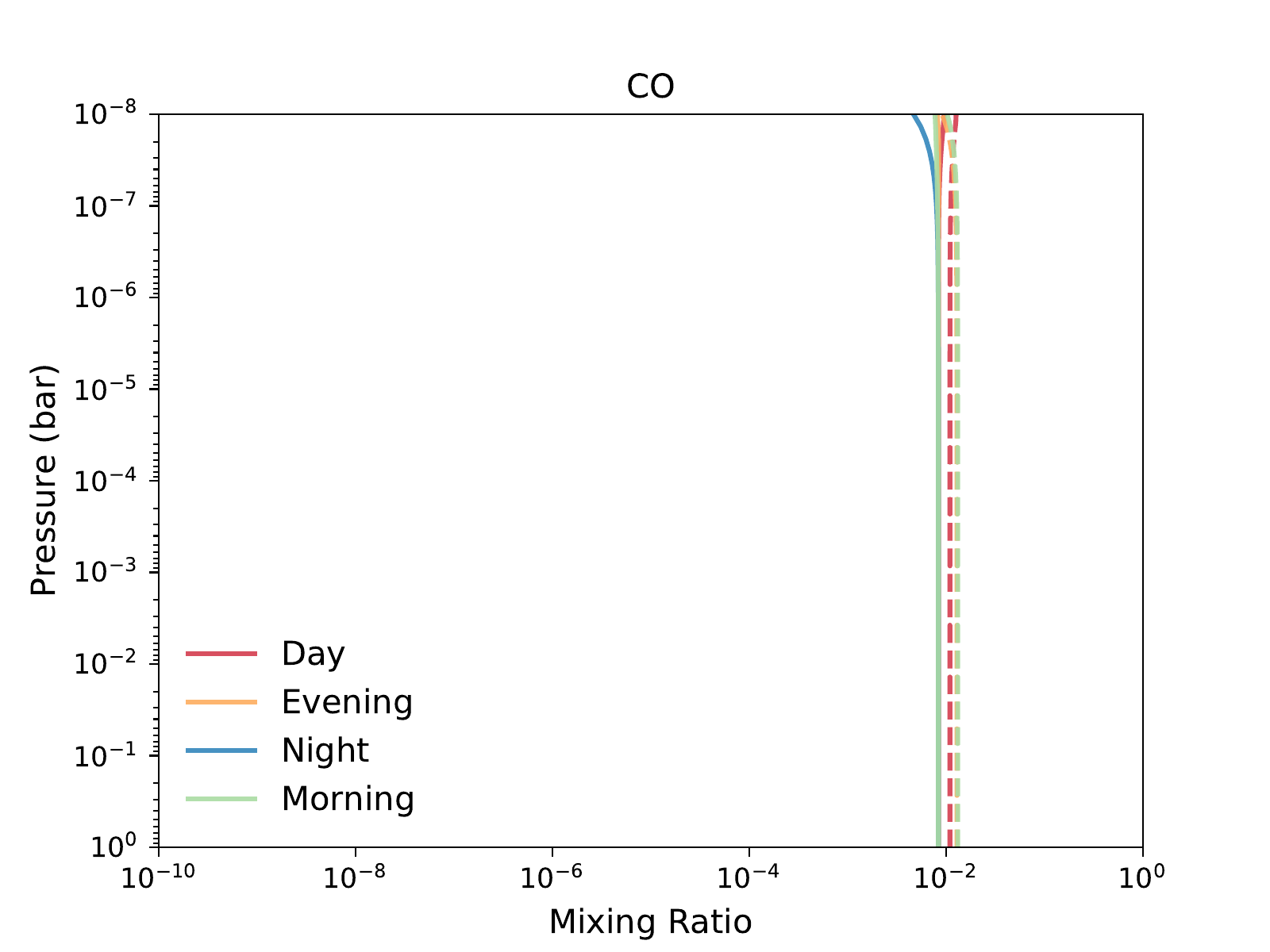}{0.5\columnwidth}{(e)}\fig{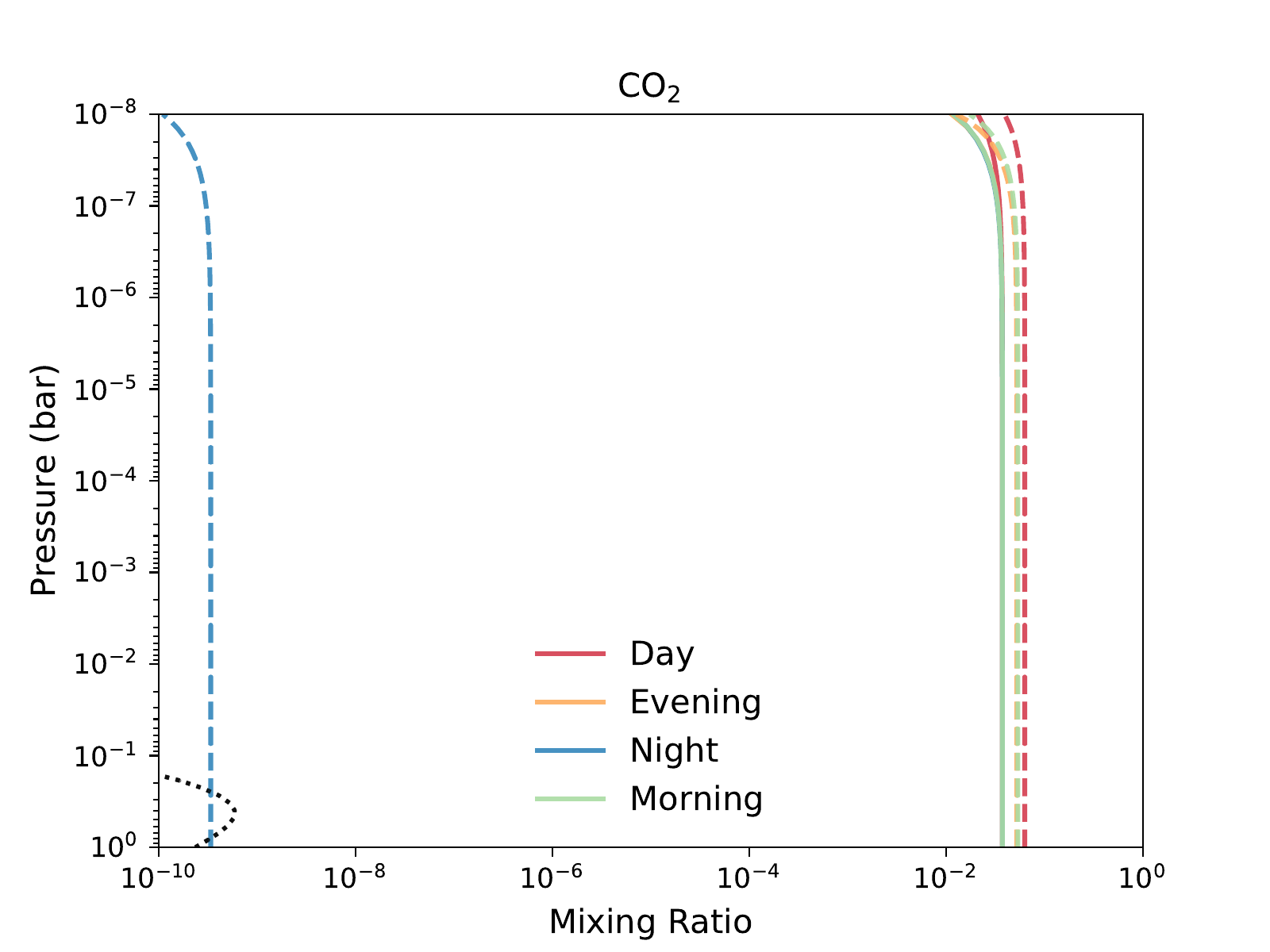}{0.5\columnwidth}{(f)}}
\end{center}
\caption{Same as Figure \ref{fig:2D} (a)--(f) except for a quiet M star.}
\label{fig:2D-5gyr}
\end{figure*}

\begin{figure*}[ht!]
\begin{center}
\gridline{\fig{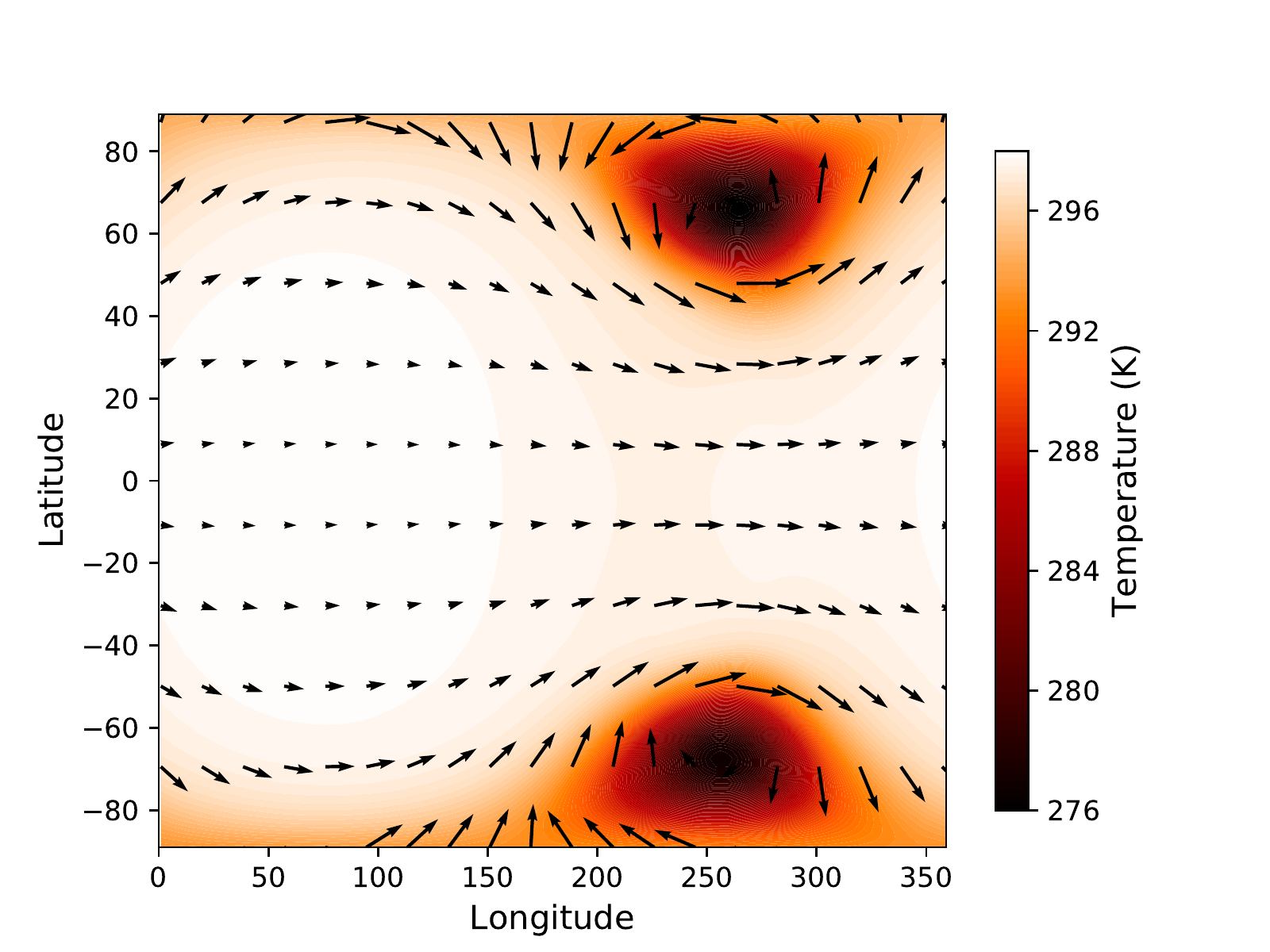}{0.55\columnwidth}{(a)}\fig{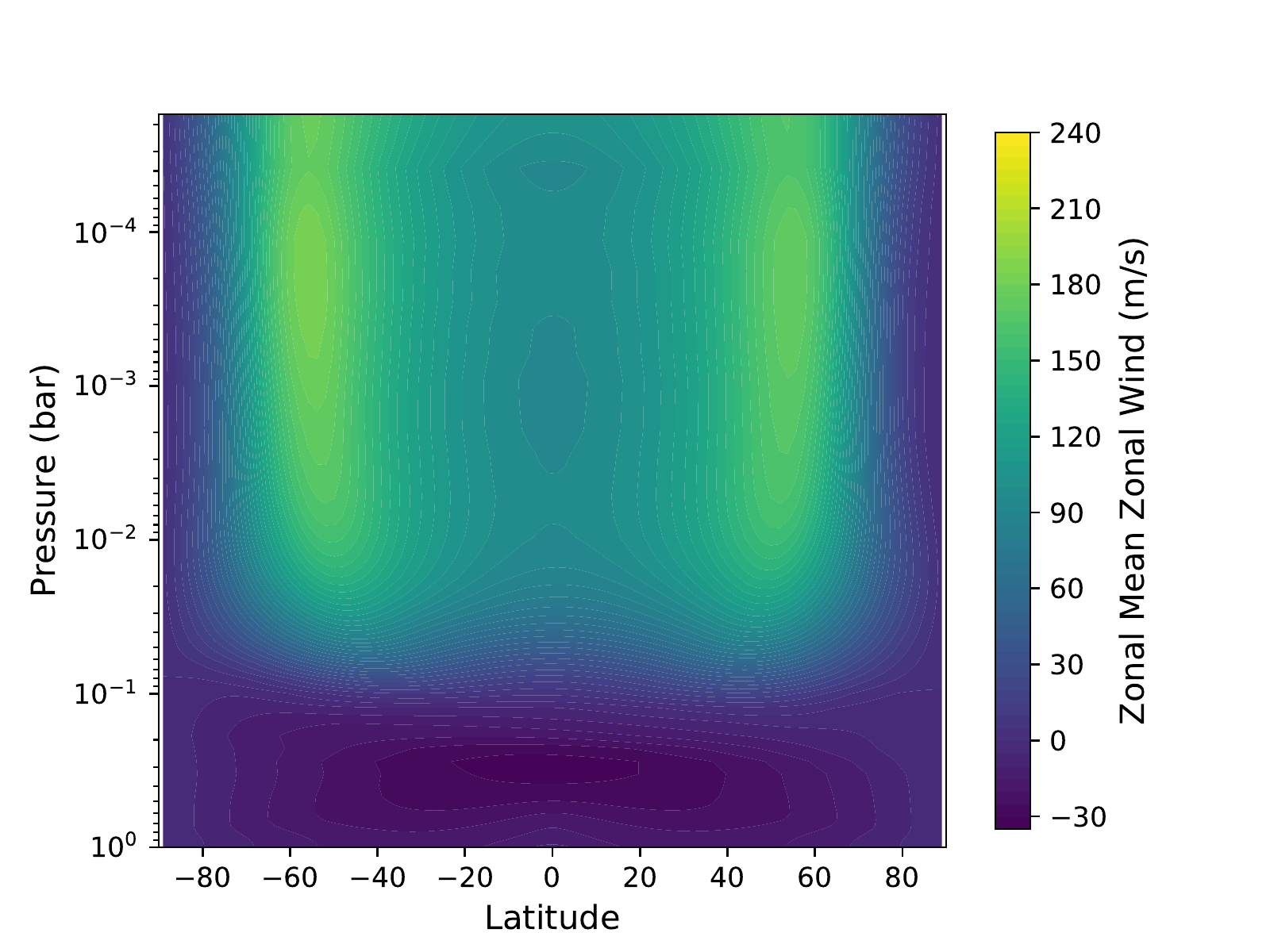}{0.55\columnwidth}{(b)}}
\gridline{\fig{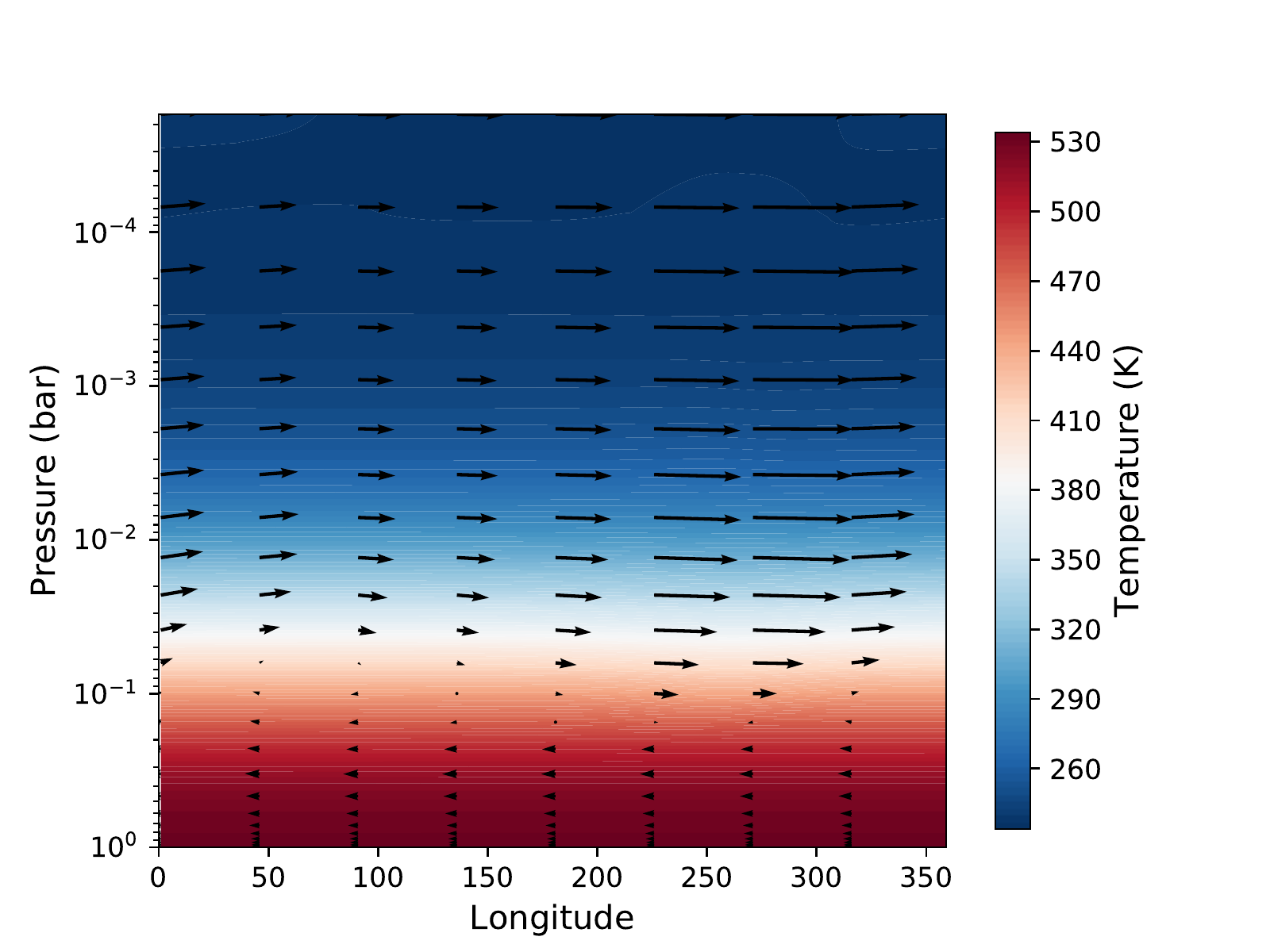}{0.55\columnwidth}{(c)}\fig{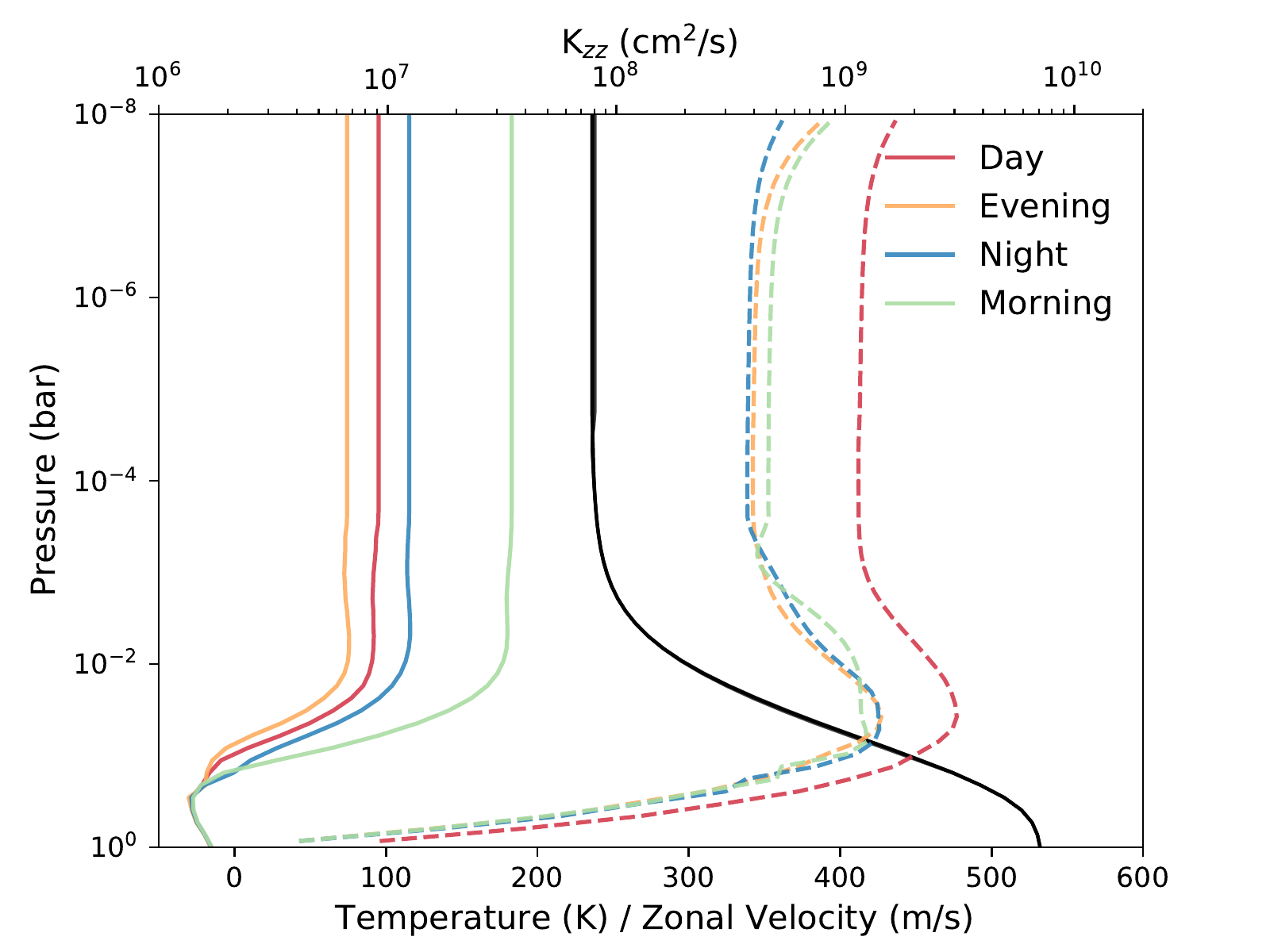}{0.47\columnwidth}{(d)}}
\end{center}
\caption{The 3D general circulation output of K2-18b with a 1-bar rocky surface, with all quantities averaged over the last 1000 (Earth) days of the 10000 day simulation. (a) shows the temperature (color scale) and horizontal wind (arrows) at 9.7 mbar level. (b) shows the zonal-mean zonal wind. (c) visualizes the temperature (color scale) and wind on the meridionally-averaged equatorial plane, with the substellar point at 0$^{\circ}$ longitude. (d) summarizes the averaged (explained in Section \ref{sec:method-2D}) temperature (black), zonal wind (solid), and eddy diffusion profiles (dashed) for the four quarters of K2-18b. All temperature profiles in black lines and overlapped because the temperature is globally uniform, whereas the zonal wind and eddy diffusion (derived from the vertical wind) profiles are color coded for the dayside, evening, nightside, and morning quarter.}
\label{fig:GCM}
\end{figure*}


\end{document}